\documentclass[a4paper]{rspublic}

\usepackage{graphicx} 
\usepackage{natbib}
\usepackage{csquotes}
\usepackage{siunitx}
\usepackage{rotating}
\usepackage{mathtools}
\usepackage{url}
\usepackage{subfigure}
\usepackage[percent]{overpic}
\usepackage{todonotes}

\renewcommand{\vec}[1]{\ensuremath{\mathbf{#1}}}
\newcommand{\unitvec}[1]{\vec{\hat{#1}}}
\newcommand{\mat}[1]{\textrm{#1}}

\newcommand{\HemeLB}{\textsc{HemeLB}}

\newcommand{\Dt}{\ensuremath{\Delta t}}
\newcommand{\Dx}{\ensuremath{\Delta x}}
\newcommand{\feq}{\ensuremath{f^{\textrm{eq}}}}
\newcommand{\fneq}{\ensuremath{f^{\textrm{neq}}}}
\newcommand{\cs}{\ensuremath{c_\textrm{s}}}

\usepackage{xcolor}
\newcommand{\rev}[1]{#1}

\title[Computational analysis of haemodynamics during
angiogenesis]{Computer simulations reveal complex distribution of
  haemodynamic forces in a mouse retina model of angiogenesis}

\author[{Bernabeu, Jones, Nielsen, et al.}]{Miguel O. Bernabeu$^{1,2}$,
  Martin Jones$^{3}$, Jens H. Nielsen$^{4}$, Timm Kr\"uger$^{5,2}$, 
  Rupert W. Nash$^{2}$, Derek Groen$^{2}$, Sebastian Schmieschek$^{2}$, 
  James Hetherington$^{4}$, Holger Gerhardt$^{3}$, Claudio A. Franco$^{6,3,*}$, 
  Peter V. Coveney$^{2,*}$}

\affiliation{ $^{1}$CoMPLEX, University College London, Physics
  Building, Gower Street,
  London, WC1E 6BT, UK\\
  $^{2}$Centre for Computational Science, Department of Chemistry,
  University College London, 20 Gordon Street, London, WC1H
  0AJ, UK\\
  $^{3}$Vascular Biology Laboratory, London Research Institute -
  Cancer Research UK, Lincoln's Inn Laboratories, 44 Lincoln's Inn
  Fields, London WC2A
  3LY, UK\\
  $^{4}$Research Software Development Team, Research Computing and
  Facilitating Services, University College London, Podium Building -
  1st Floor, Gower Street, London, WC1E 6BT, UK\\
  $^{5}$Institute for Materials and Processes, University of
  Edinburgh, King's Buildings, Mayfield Road, Edinburgh EH9 3JL,
  Scotland, UK\\
  $^{6}$Instituto de Medicina Molecular, Faculdade de Medicina Universidade de Lisboa, Lisboa 1649-028, Portugal\\
  $^{*}$Equally contributing senior authors}

\begin{document}

\maketitle

\begin{abstract}{angiogenesis, mouse, retina, blood flow, simulation,
    shear stress, image-based, lattice-Boltzmann}
  There is currently limited understanding of the role played by
  haemodynamic forces on the processes governing vascular
  development. One of many obstacles to be overcome is being able to
  measure those forces, at the required resolution level, on vessels
  only a few micrometres thick. In the current paper, we present an
  \emph{in silico} method for the computation of the haemodynamic
  forces experienced by murine retinal vasculature (a widely used
  vascular development animal model) beyond what is measurable
  experimentally. Our results show that it is possible to reconstruct
  high-resolution three-dimensional geometrical models directly from
  samples of retinal vasculature and that the lattice-Boltzmann
  algorithm can be used to obtain accurate estimates of the
  haemodynamics in these domains. \rev{We generate flow models from 
  samples obtained at postnatal days (P) 5 and 6. 
  Our simulations show important differences between the flow patterns 
  recovered in both cases, including observations of regression 
  occurring in areas where wall shear stress gradients exist. We propose two possible
  mechanisms to account for the observed increase in velocity and wall
  shear stress between P5 and P6: i) the measured reduction in 
  typical vessel diameter between both time points, ii) the reduction 
  in network density triggered by the pruning process.}
  % Our findings show that the flow
  % patterns recovered are complex, that branches of predominant flow
  % exist from early development stages, and that the pruning process
  % tends to make the wall shear stress experienced by the capillaries
  % increase by removing redundant segments. 
  The
  methodology developed herein is applicable to other biomedical
  domains where microvasculature can be imaged but experimental flow
  measurements are unavailable or difficult to obtain.
\end{abstract}

% \setcounter{section}{-1}
% \section{Notes}

% Three main topics: i) describe how we turn plexus images into 3D
% models suitable for blood flow simulation, ii) survey the literature
% for the relevant data to set up the simulations (typical flow rates
% and pressures, Reynolds and Womersley numbers, etc.)  iii) show that
% our numerical method (\emph{i.e.} lattice-Boltzmann (LB)) is accurate
% enough for the simulation of wall shear stress in sparse geometries at
% the required Reynolds/Womersley number and flow characteristics
% (steady \emph{vs.}  pulsatile flow). Finish with a few shear stress
% maps in one of the plexus geometries. No in depth analysis or linking
% with biology required, just proof of concept simulations. \todo{Plots
%   will be made consistent and beautified in due course.}

%latex2rtf plexus_flow_methods_paper_manuscript.tex 

\section{Introduction}

Despite recent advances in vascular biology, the mechanisms
underpinning vascular development remain poorly understood. It is
therefore crucial to gain further insight into the mechanisms
governing the formation of complex vascular networks and their response to external
stimuli. The translation of these results holds the key to the
improvement of therapies modulating vascular patterning and sprouting
for the treatment of stroke, ischaemia, retinopathies or cancer, the
leading cause of death worldwide.

\begin{figure} \centering
\begin{tabular}{cc}
\includegraphics{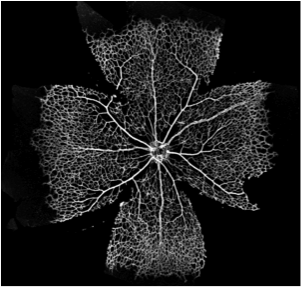} & \includegraphics{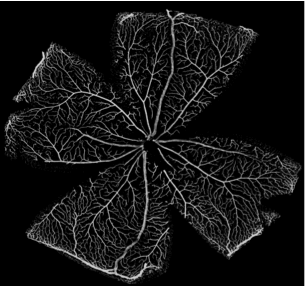}
\end{tabular}
\caption{Murine retinal vascular plexus 6 and 21 days postnatal (left
  and right panel, respectively). Within days, the primitive vessel
  network remodels into mature vasculature. Samples were collected,
  mounted, and imaged as described in Section \ref{sec:methods}.}
\label{fig:retinas}
\end{figure}

One of the pressing questions in the field is establishing how
primitive vessel networks remodel into a hierarchically branched and
functionally perfused network of arteries, arterioles, capillaries,
and venules (Figure \ref{fig:retinas}). In recent years, the main molecular mechanisms
regulating endothelial cell behaviour during vessel formation have
been elucidated using experimental techniques
\citep{Jones2006,Potente2011}. However, important challenges remain:
i) understanding how cell-level mechanisms integrate to give rise to
systems-level behaviour and, ii) understanding the impact in vascular
patterning of the interplay between cellular molecular regulation and
haemodynamic forces (\emph{i.e.} vascular mechanotransduction).
These problems are hard to address due to the multiscale and multiphysics
nature of the processes involved. Systems-level behaviour arises from
highly non-linear, tightly coupled interactions between subprocesses
at different spatial and temporal scales.
Furthermore, it has been recently proposed \citep{Bentley2013} that a
tighter integration between experimental and computational work is
required in order to tackle these questions. Working in a feedback
loop, computational models should be capable of generating new
hypotheses, rather than merely reproducing experimental
data. In turn, experiments should provide new biological insights based
on these hypotheses and help to further refine computational models.
%In the words of Denis Noble, \textquote[\citet{Noble2002}]{Biology is
%  set to become highly quantitative in the 21st century. It will
%  become a computer-intensive discipline.}

%
% TEXT ABOUT FSG MODELLING (IN CASE IT'S REQUIRED)
%
% In recent years, the biomechanics community has shown an increasing
% interest in the integration of fluid-solid interaction (FSI) models
% with models of tissue growth and remodelling (G\&R) into so-called
% fluid-solid-growth (FSG) models \cite{Humphrey2008}. More precisely,
% FSI computations are performed over one or more cardiac cycles and the
% resulting distribution of haemodynamic forces is used to inform models
% of G\&R simulated over periods of days or weeks. The predicted changes
% in the structure of the vessels are fed back into the FSI model and
% the system is iterated in order to study a process of interest
% (e.g. vessel pruning during angiogenesis). To date this modelling
% approach has been successfully applied to the study of intracranial
% and aortic aneurysms. To the best of our knowledge, it has never been
% applied to the study of developmental vascular remodelling.

Multiple animal models have been proposed for the study of vascular
development. Examples include the mouse retinal and embryonic vasculature
\citep{Geudens2011}, zebrafish vasculature
\citep{Lawson2002,Siekmann2007}, and hyaloid vasculature
\citep{Lobov2005}.  In recent years, there has been increasing
interest in the development of \emph{in silico} models for the close
inspection of certain vascular developmental aspects. To date, most work
concerning simulation of retinal haemodynamics for the study of
vascular mechanotransduction (see Section
\ref{sec:lit_rev}\ref{sec:previous} for a review) has suffered from a number of
limitations including: i) limited availability of  spatial information due to the use of low
resolution imaging modalities, ii) oversimplification of the
haemodynamics by considering the retinal plexus to be a network of
one-dimensional vessel segments, and iii) unavailability of the
computer code developed. We believe that
the model simplifications cited, although appropriate in some applications, may
fail to capture complex flow patterns important for understanding the
interplay between molecular regulation and haemodynamics during
development. Hence, in the current work we introduce a computational
workflow --- and make the source code available --- aimed at
generating \emph{in silico} estimates of the haemodynamic forces
acting on samples of mouse retinal vasculature imaged during
development, typically within the first postnatal week.
%Sample preparation and imaging have been previously described in \citet{Franco2013}.  
The workflow involves the following steps. First, high-resolution
scanning confocal microscope images are obtained and segmented in
order to generate a binary mask of the vessel lumen. Second, luminal
centrelines and radii are computed in a process known as
skeletonisation. Next, three-dimensional (3D) models of the luminal surface are
reconstructed based on the computed
skeleton. %; assuming circular cross-section.
Finally blood flow simulations are run in order to obtain estimates of
blood velocity and wall shear stress with an open source
highly-parallel computational fluid dynamics (CFD) solver, known as
\HemeLB{} \citep{Mazzeo2008}.

The purpose of this paper is therefore three-fold. First, to
describe the computational methods developed and to survey the
literature for data not accessible in our experiments but necessary
for model set-up. Second, to validate our methods in simplified
scenarios where analytical solutions are known. Third, to present and
to analyse simulations in order to gain insight into
the dynamics of retinal blood flow during development. The 
paper is structured as follows. In Section \ref{sec:lit_rev}, we survey the
literature for previously proposed models of retinal flow
and for the experimental data necessary to set up our
simulations. Next, in Section \ref{sec:methods} we present the methods
used for image processing and 3D model reconstruction as well as the
validation methodology adopted. Section \ref{sec:results} presents the
main results on model reconstruction, validation, and a set of simulations on the reconstructed 3D models. Finally,
Section \ref{sec:conclusions} summarises the main contributions of the
work and outlines the areas where we plan to apply the
computational pipeline developed.

\section{Retinal vascular structure and flow}
\label{sec:lit_rev}
\subsection{Vascular structure and its development}

Angiogenesis defines the formation of new blood vessels from
pre-existing ones and can be split into two distinct phases: sprouting
and remodelling. During sprouting, new vessels form and invade
avascular ischaemic areas, where tissues experience hypoxia and
nutrient deprivation. This process is modulated by the secretion of
various growth factors, including vascular endothelial growth factor
(VEGF), through a cascade of signalling events. The end point of
this phase is the formation of a highly branched and poorly perfused
network of capillary connections. Remodelling is responsible for the
creation of a hierarchically branched and efficient vascular tree, containing
defined arteries and veins and an optimised vascular capillary
network. A vital step during vascular remodelling is the removal of
redundant vessel segments: vessel pruning. Importantly, angiogenesis
is a very dynamic process occurring not only during development but
also in adulthood (\emph{e.g.} wound healing and tumour formation).

The neonatal mouse retina has become one of the main experimental
models for the study of the mechanisms involved in blood vessel
development and patterning \citep{Fruttiger2007,Uemura2006,
  Gariano2005}.  The mouse retina is avascular at birth and develops
through a consistent series of events. Astrocytice (a type of glial
cell) and neuron derived vascular endothelial growth factor A (VEGFA)
stimulates sprouting angiogenesis from pre-existing blood vessels at
the optic nerve. Under a gradient of VEGFA, vessels expand radially in
the superficial layer of the retina, in a very characteristic pattern
\citep{Potente2011}. It takes this process around 8 days to 
cover the entire surface of the mouse retina. Vascularisation of the
superficial layer is followed by a second phase of sprouting, where
endothelial cells from the superficial venous plexus sprout and
penetrate the deeper layers of retina to form, firstly, a deep and,
secondly, an intermediate capillary bed \citep{Brown2005}. The
vascular plexus finally matures about 20 days after birth.

In the mature mouse retina, vessels are found to be predominantly
arteriolar in the superficial layer and predominantly venular in the
deep capillary bed \citep{Paques2003}. The artery feeding the retina
arrives at the optic disc and divides into \numrange{8}{9} radiating
retinal arteries. These arteries, with luminal diameter of up to
\SI{28}{\micro\metre} \citep{Ninomiya2006}, side branch into smaller
arterioles at close to 90 degrees angles from the parent arteries. The
arterioles (\SIrange{10}{12}{\micro\metre} in diameter
\citep{Ninomiya2006}) take a relative long course before abruptly
changing direction to run towards the intermediate and subsequently
deep capillary beds (\SIrange{5}{6}{\micro\metre} in diameter
\citep{Ninomiya2006}). Some authors (\emph{e.g.}  \citet{Paques2003})
have suggested that the superficial layer is mostly capillary-free
with only a few direct connections between arteries and veins.

\subsection{Haemodynamics}
\label{sec:ret_haem}

% Survey the literature for experimentally observed parameters to be
% used to set up the simulations. This includes:
% \begin{itemize}
% \item Typical values of pressure and flow rate to be used at the
%   inlets and outlets.
% \item Whether flow can be assumed steady or it's pulsatile.
% \item Explicit computation of Reynolds and Womersley numbers.
% \end{itemize}

The relationship between haemodynamics and pathogenesis of various eye
disorders has prompted researchers to analyse retinal blood flow in
both basic and clinical research domains. Early examples are the work
of \citet{Feke1989}, who measured total retinal blood flow and its
regional distribution in humans, and \citet{Alm1973} who studied blood
flow rates in various tissues of the primate eye. Later advances in
imaging techniques (\emph{e.g.} optical coherence tomography and
related modalities, laser Doppler velocimetry)
% such as ultrasound biomicroscopy (UBM) \cite{addcite}, optical
% coherence tomography (OCT) \citep{Huang1991}, and related OCT
% modalities
have expanded our understanding of retinal haemodynamics. High
resolution \emph{in vivo} measurements of retinal flow have been
obtained in various species: mouse \citep{Brown2005}, rat
\citep{Zhi2011}, and human \citep{White2003, Wang2009}. Several
authors \citep{White2003, Yazdanfar2003} presented evidence of the
pulsatile nature of retinal blood flow despite early claims
\citep{Alm1973} that only retinal arteries --- and not veins ---
exhibit systolic to diastolic flow rate variations. Further
developments enabled quantitative analysis of typical arterial and
venous flow in both healthy \citep{Wang2007} and diseased
\citep{Wang2009} human retinas as well as during development in mice
\citep{Brown2005}.  Table \ref{ta:vels_flows} compiles some of the
measurements of murine and human retinal vessel diameter, velocity,
and flow rate available in the literature.
\begin{sidewaystable}
  \caption{Experimentally observed values of vessel diameter and flow
    rate in different parts of the murine and human adult retina as
    reported in various publications. BTBR and C57BL/6J are two common
    mouse strains used as models of human disease. Values given as
    mean $\pm$ standard error of the mean, when possible. The values 
    reported by \citet{Zhi2012} are measured at five different arteries/veins
    and averaged over three independent measurements. Significant
    inconsistencies are found across the surveyed literature: a)
    \citet{Wright2012} and \citet{Wang2007} measured flow rates
    approximately one order of magnitude higher than \citet{Zhi2012},
    b) the arterial diameter measured in \citet{Ninomiya2006} is
    substantially lower than in \citet{Wright2012}.}
\label{ta:vels_flows}
\centering
\begin{tabular}{lllp{4cm}}
  \\ \hline
  Authors & Species & Measurement & Value \\ \hline
  \citet{Wright2012} & 30-week-old C57BL/6J male mice & Arterial
  diameter & \SI{\sim57}{\micro\metre} \\
  & & Venous diameter & \SI{\sim62}{\micro\metre} \\
  & & Mean arterial velocity &
  \SI{\sim25}{\milli\metre\per\second}\\
  & & Mean venous velocity & \SI{\sim24}{\milli\metre\per\second}\\
  & & Arterial flow rate & \SI{\sim3.9}{\micro\litre\per\minute} \\ %\SI{\sim65}{\nano\litre\per\second} \\
  & & Venous flow rate & \SI{\sim4.8}{\micro\litre\per\minute} \\ %\SI{\sim80}{\nano\litre\per\second} \\
  \hline
  \citet{Zhi2012} & 22-week-old BTBR female mice & Arterial flow rate &
  \SIlist{0.40\pm0.04;0.55\pm0.06;0.50\pm0.05;0.48\pm0.05;0.40\pm0.04;0.45\pm0.06}{\micro\litre\per\minute} \\ 
  & & Venous flow rate &
  \SIlist{0.45\pm0.04;0.62\pm0.06;0.60\pm0.04;0.59\pm0.05;0.38\pm0.03;0.63\pm0.06}{\micro\litre\per\minute} \\ 
  \hline
  \citet{Wright2009} & 11--12-week-old C57BL/6 male mice & Arterial diameter & \SI{60.4\pm0.7}{\micro\metre} \\
  & & Venous diameter & \SI{69.3\pm1.3}{\micro\metre} \\
  & & Mean arterial velocity &
  \SI{28.3\pm1.4}{\milli\metre\per\second}\\
  & & Mean venous velocity & \SI{26.3\pm1.2}{\milli\metre\per\second}\\
  \hline
  \citet{Ninomiya2006} & Four-month-old mice (\emph{ex vivo}) &
  Arterial diameter & up to  \SI{28}{\micro\metre} \\
 & & Capillary diameter & \SIrange{5}{6}{\micro\metre} \\
  \hline
\citet{Wang2007} & Adult human & Arterial diameter &
\SI{91.23\pm11.80}{\micro\metre} \\
 & & Arterial peak velocity &
 \SI{24.15\pm1.50}{\milli\metre\per\second}\\
 & & Arterial flow rate &
 \SI{6.83\pm1.75}{\micro\litre\per\minute}\\
 & & Venous diameter &
\SI{69.83\pm3.52}{\micro\metre} \\
 & & Venous peak velocity &
 \SI{46.43\pm1.42}{\milli\metre\per\second}\\
 & & Venous flow rate &
 \SI{6.42\pm0.72}{\micro\litre\per\minute}\\
\end{tabular}
\end{sidewaystable}

Of relevance to our study is the work by \citet{Brown2005}, who
obtained \emph{in vivo} measurements of blood flow in various parts of
the murine eye including the retina from birth to postnatal day (P) 16. Table
\ref{ta:brown_vels} presents some of their findings. A clear trend of
increase in retinal blood flow after P3 is observed. The authors
attribute the large variability of the results (note standard
deviation in Table \ref{ta:brown_vels}) to the natural variation in
the time course of the remodelling processes involved.
\begin{table}
  \caption{Retinal peak velocities measured in CD-1 mice by
    \citet{Brown2005}. $\sigma$: standard deviation, N: number of samples.}
  \label{ta:brown_vels}
  \centering
  \begin{tabular}{lccc}
    \\ \hline
    Age & Velocity (\SI{}{\cm\per\second}) & $\sigma$
    (\SI{}{\cm\per\second}) & N \\ \hline
    P0 & 0.32 & 0.09 & 5 \\
    P1 & 0.71 & 0.29 & 7 \\
    P2 & 1.22 & 0.29 & 6 \\
    P3 & 0.58 & 0.07 & 6 \\
    P4 & 4.62 & 1.09 & 5 \\
    P5 & 2.70 & 0.94 & 5 \\
    P6 & 5.26 & 1.79 & 5 \\
    P7 & 4.30 & 0.07 & 6 \\
    P8 & 3.53 & 1.73 & 7 \\
    P10 & 2.79 & 0.14 & 5 \\
    P12 & 4.35 & 0.52 & 5 \\ \hline
  \end{tabular}
\end{table}

Several authors have also studied the typical pressure difference
driving flow in the retina, the so-called ocular perfusion pressure
(OPP). The pressure at the central retinal artery is often
approximated with mean arterial pressure (MAP) measurements at eye
level (\emph{e.g.} carotid arterial pressure \citep{Wright2008},
subclavian artery \citep{Hardy1994}). The effective venous pressure is
considered equivalent to the intraocular pressure (IOP)
\citep{Kiel2010}. Table \ref{ta:map_iop} summarises the values of MAP
and IOP reported by several authors. OPP values of
$\sim$\SI{57}{\mmHg} are consistently reported across
species. Finally, retinal blood flow is known to be autoregulated by
the modulation of retinal vessel compliance in response to MAP and IOP
variations \citep{Paques2003, Hardy1994}.
\begin{table}
  \caption{Values of mean arterial pressure (MAP) and intraocular
    pressure (IOP) reported in the literature for different species.}
\label{ta:map_iop}
\centering
\begin{tabular}{lccc}
\\ \hline
Authors & Species &  MAP (\SI{}{\mmHg}) & IOP (\SI{}{\mmHg}) \\ \hline
\citet{Wright2008} & 16-week-old mice & 68.2 $\pm$ 2.0 & 11.6 $\pm$ 0.4 \\
\citet{Hardy1994} & 1--3-day-old piglets & 70 $\pm$ 6 & 13 $\pm$ 2 \\ \hline
\end{tabular}
\end{table}

From a rheological point of view, blood is a shear-thinning fluid
(\emph{i.e.} its viscosity is a decreasing function of shear rate
\citep{Chien1970}). When flowing at sufficiently large shear rates
(typically greater than \SI{\sim1000}{\per \second}) blood can be
modelled as a Newtonian fluid (\emph{i.e.} constant viscosity) with no
significant effect on the simulated haemodynamics (see
\citet{Bernabeu2013Rheology} and references therein). However, at lower shear
rates, viscosity quickly increases due to \emph{e.g.} red blood cell
(RBC) aggregation. Our work concerns simulation of blood flood in
small arterioles, venules, and capillaries where shear rate is
expected to be lower than the aforementioned threshold. In fact,
\citet{Nagaoka2006} measured shear rates as low as
\SI{606\pm115}{\per\second} in the venules of the human retina. Therefore, we
will take the shear-thinning properties of blood into account in order
to improve the fidelity of the shear stress computed in our
model. Table \ref{ta:visc_values} compiles some of the values of
murine blood viscosity as a function of shear rate available in the
literature. Other rheological properties derived from the presence of
RBCs, such as the F\aa hr\ae us-Lindqvist effect (see \emph{e.g.} \citep{Pries1996}),
will not be considered.
\begin{table}
\caption{Blood viscosity as function of shear rate in mice.}
\label{ta:visc_values}
\centering
\begin{tabular}{llcc}
  \\ \hline
  Authors & Animals & Shear rate (\SI{}{\per\second}) &  Viscosity (\SI{}{\milli\Pa\second})  \\ \hline
  \citet{Vogel2003}& 4--7-month-old C57Bl/6 mice & 2 &  18.94
  (average, $n=11$)\\
  &  & 5 & 13.33 \\
  &  & 11 & 10.52 \\
  &  & 23 & 8.07 \\
  &  & 45 & 6.31 \\
  &  & 90 & 5.96 \\
  &  & 225 & 4.91 \\
  &  & 450 & 3.85 \\ \hline
  \citet{Windberger2003} & 4--8-month-old BALB/c mice & 0.7 & 13.36 (median, $n=37$) \\
  &  & 2.4 & 10.56 \\
  &  & 94 & 4.87 \\ \hline
\end{tabular}
\end{table}

Finally, we note that experimental data on haemorheology changes during development is
limited. \citet{Windberger2005} observed a steady increase in blood
viscosity in rabbits and cats from fetal stages to
adulthood. The changes were more pronounced within lower
shear rates regimes: from \SI{3.00}{\milli\Pa\second} to
\SI{9.29}{\milli\Pa\second} at \SI{0.7}{\per\second} and from
\SI{2.48}{\milli\Pa\second} to \SI{3.62}{\milli\Pa\second} at
\SI{94}{\per\second} during the first 30 days of life in rabbits. Due
to the scarcity of available data, we will not include this effect
in our model.

% 2, 18.94736842105263
% 5,13.33333333333334
% 11,10.526315789473706
% 23,8.07017543859648
% 45,6.3157894736842
% 90,5.9649122807017
% 225,4.912280701754414
% 450,3.85964912280701

% mean, n=37
% 0.7, 13.367
% 2.4, 10.563
% 94, 4.879

The experimental data on pressure distributions and haemorheological
properties surveyed in this section will be used to set up our flow
simulations. This is done due to the impossibility of obtaining such
information directly from our experimental model. The data summarised
in Table \ref{ta:vels_flows} will be used to validate our experimental
measurements of vessel diameter and \emph{in silico} estimates of blood
velocity and flow rate. We now turn our attention to previously
proposed models of retinal blood flow.

\subsection{Previous modelling and simulation studies}
\label{sec:previous}

In one of the earliest works on retinal haemodynamics modelling and
simulation, \citet{Ganesan2010} state that ``although a relatively
good understanding of the retinal anatomy and vascular network has
been developed through extensive studies [...] there is a complete
lack of numerical modeling of retinal circulation in the
literature''. In the same study, an image-based network model of the
retinal vasculature was developed. The location and length of
non-capillary vessel segments was extracted from confocal microscopy
images of flat-mounted mouse retinas and a rule-based network model
used to approximate the structure of the capillary bed. The
haemodynamics were greatly simplified by considering vessel segments
to be straight with piece-wise constant radius and flow to be laminar
(\emph{i.e.} a 1D network model), therefore neglecting complex fluid
patterns that may appear in curved vessels even at low Reynolds numbers
\citep{Siggers2005}. Their results show that wall shear stress in the
capillaries stays mainly in the \SIrange{4}{11}{\pascal} range with
values as high as \SI{20}{\pascal}. These magnitudes are substantially
higher and with a much greater spread than those reported in the main
retinal arteries and veins. Prior to this work, \citet{Liu2009} also
developed an image-based retinal flow model for the study of oxygen
transportation in the retina. In this case, only a subset of the
retinal vasculature (\emph{i.e.} an artery and a number of branching
arterioles) was reconstructed from a healthy human fundus camera
image. The two-dimensional steady-state Navier-Stokes equations were
solved in the domain. The model was used to predict pressure drops and
oxygen saturation distribution.

More recently, \citet{Chen2012} developed a mathematical model of
blood flow in the zebrafish larvae midbrain vasculature (another
typical model for the study of vascular development). The morphology
of the vessel network was recovered from \emph{in vivo} images at
different developmental stages. Haemodynamics were also modelled using
a laminar flow in straight circular pipe simplification. Both steady
and pulsatile flow were compared with little difference in overall
dynamics. The flow model was in turn coupled to a phenomenological
model of changes in vessel diameter as a function of shear stress
(without any explicit mechanism of endothelial cell migration). The
resulting coupled model was used to predict vessel pruning in several
zebrafish larvae midbrain vasculature samples with a reported 75\%
accuracy. These results support the hypothesis that wall shear stress
is a major factor in vessel pruning during angiogenesis.

Finally, \citet{Watson2012} developed a comprehensive model of murine
retinal angiogenesis including cell migration during sprouting, blood
flow, oxygen distribution, and the main chemotactic gradients involved
in vessel development and pruning. In their work, vascular pruning was
mainly driven by the downregulation of growth factors but no explicit
mechanobiological mechanisms were considered. Blood flow simulation was
also performed based on the 1D network simplification described above.

The articles cited in this section demonstrate increasing interest
in the modelling and simulation of retinal
haemodynamics. Computational models have been developed for human,
mouse, and zebrafish retinal vasculature. A common application is the
study of vascular development dynamics (angiogenesis in
particular). In our opinion, the previous works, although seminal,
share one or more of the following limitations: i) the choice of
imaging modality only allows the recovery of a subset of the retinal
vasculature, ii) flow dynamics are greatly simplified by the use of 1D or
2D approximations, and iii) the computer code developed is, to the
best of our knowledge, not freely available.  In the current work, we
aim at developing an open source computational workflow for the
generation of high resolution estimates of the haemodynamic forces
experienced by murine retinal vasculature based on confocal microscope
images. The following section describes the methodology employed.

%\citep{Harrison2011} Purtscher Retinopathy: An Alternative Etiology Supported by Computer Fluid Dynamic Simulations

\section{Methods}
\label{sec:methods}

\subsection{Image processing and 3D model reconstruction}
\label{sec:image_proc}

% We want to describe how we reconstruct 3D models from images. This
% involves talking about:
% \begin{itemize}
% \item Basic description of the experimental setup and image
% acquisition.
% \item Martin's algorithms for segmenting the images (\emph{i.e.}
% creating binary masks) and computing image skeletonisation and radii.
% \item Miguel's code to turn the skeletonisation into 3D models (credit
% VTK and VMTK libraries).
% \end{itemize}

The preparation of retinal vascular plexus samples for imaging and
analysis has been previously described in \citet{Franco2008}. Briefly,
plexus samples were collected from \rev{five and six}-day-old
wildtype mouse pups and fixed with 2\% paraformaldehyde in phosphate
buffered saline (PBS) for 5 hours at $4^\circ$C, thereafter retinas
were dissected in PBS. Blocking/permeabilisation was performed using
Claudio's blocking buffer (CBB), consisting of 1\% FBS (Gibco), 3\%
BSA (Sigma), 0.5\% triton X100 (Sigma), 0.01\% Na deoxycholate
(Sigma), 0,02\% Na Azide (Sigma) in PBS pH=7.4 for 2-4 hours at
$4^\circ$C on a rocking platform.  Samples were stained with
endothelial luminal marker (ICAM2) and incubated at the desired
concentration in 1:1 CBB:PBS at $4^\circ$C overnight in a rocking
platform. Finally, retinas were mounted on slides using Vectashield
mounting medium (Vector Labs, H-1000) and imaged with a Carl Zeiss
LSM780 scanning confocal microscope (Zeiss).  Manual preprocessing of
the plexus image was performed with Photoshop CS5 (Adobe) in order to
remove major imaging artefacts, allowing a simple thresholding method
to be used to produce a binary image of the entire retinal plexus
segment. The binary image was skeletonised using a MATLAB (The
MathWorks, Inc.)
interface\footnote{\url{http://www.mathworks.co.uk/matlabcentral/fileexchange/27543-skeletonization-using-voronoi}}
to the Voronoi tessellation algorithm implemented in the QHull library
\citep{Barber1996}. The radius at each Voronoi vertex was calculated
from the maximum inscribed circle in two-dimensions.

Based on the image skeleton and computed radii, a three-dimensional
triangulation of the plexus luminal surface was generated by assuming
vessel circular cross section with the VTK \citep{Schroeder2003} and
VMTK \citep{Antiga2008} libraries.  This simplification is made based
on our own histological analysis and on the observations of
\citet{Feke1989}, who cite histological evidence of retinal arteries
being circular in cross-section while retinal veins exhibit a higher
tendency towards flattening. The volume contained within the surface
was discretised as a regular grid in order to generate the
computational domain necessary for simulation (see next section for details). All the scripts
developed are freely available at
\url{https://github.com/UCL/BernabeuInterface2014}.

\subsection{Simulation setup}
\label{sec:sim_setup}
\rev{Let $\Omega$ be a three-dimensional domain with boundary $\partial
\Omega$.} The CFD package \HemeLB{} \citep{Mazzeo2008} was used to solve
numerically the Navier Stokes equations for generalised Newtonian
incompressible fluids. \rev{For $\vec{x} \in \bar\Omega$ and time $t \in [0,\mathcal{T}]$:}
\begin{align}
  \nabla \cdot \vec{v} &= 0\;, \label{eq:ns_mass}\\
  \rho\left(\frac{\partial \vec{v}}{\partial t} + \vec{v}\cdot\nabla\vec{v}\right) &= -\vec{\nabla}P +
  \vec{\nabla}\cdot\mat{T}\;, \label{eq:ns_momentum}
\end{align}
where $\rho$ is the density, $\vec{v}(\vec{x},t)$ is the velocity vector,
$P(\vec{x},t)$ is the pressure, $\mat{T}(\vec{x},t)$ is the deviatoric part of
the stress tensor
\begin{align}
  \mat{T}_{i j} & = 2\eta \mat{S}_{i j}\;, \\
  \mat{S}_{i j} &=\frac{1}{2}\left(\frac{\partial v_j}{\partial
    x_i}+\frac{\partial v_i}{\partial x_j}\right)\;,
\end{align}
and $\eta$ is the dynamic viscosity which may depend on the shear
rate $\dot{\gamma}$, \emph{i.e.} $\eta(\dot{\gamma})$,
\begin{align}
  \dot{\gamma} &= \sqrt{2 S_{ij}S_{ij}}\;,
\end{align}
where $i,j=1,2,3$ and summation over repeated indices is assumed. Note 
that in the case of Newtonian fluids, $\eta(\dot{\gamma}) = \eta = \textrm{const}$. 

\rev{Let $\partial \Omega_{w,i,o}$ be the wall, inlet, and outlet portions of
the domain boundary, respectively, such that $\partial \Omega = \partial
\Omega_w \cup \partial \Omega_i \cup \partial \Omega_o$. Equations
\eqref{eq:ns_mass}--\eqref{eq:ns_momentum} are closed with the following
initial condition}
\begin{align}
  \vec{v}(\vec{x},0 ) = \vec{0}\;,\;\;\;\;\vec{x} \in \bar\Omega\;,
\end{align}
\rev{and boundary conditions}
\begin{align}
      \vec{v} &= \vec{0}\;, &\vec{x} &\in \partial \Omega_w\;, \\
      P\unitvec{n} - \frac{\eta}{\rho} \nabla \vec{v} \cdot \unitvec{n} &= P_i \unitvec{n}\;, &\vec{x} &\in \partial \Omega_i\;,\\
      P\unitvec{n} - \frac{\eta}{\rho} \nabla \vec{v} \cdot \unitvec{n} &= P_o \unitvec{n}\;, &\vec{x} &\in \partial \Omega_o\;,
\end{align}
\rev{(\emph{i.e.} a pressure drop problem as formulated in \citet{Heywood1996,FormaggiaBook})
where $\unitvec{n}(\vec{x})$, $\vec{x}\in \partial \Omega$, is the boundary 
normal vector and $P_{i,o}(t)$ are the pressures at the inlet and outlet, respectively.}
\HemeLB{} uses the lattice-Boltzmann (LB) algorithm
(see \ref{ap:lb} for a brief introduction) and runs efficiently on large scale high performance
computing resources \citep{Groen2013Performance}. \HemeLB{}'s source code is
available under LGPL licence and can be downloaded from
\url{http://ccs.chem.ucl.ac.uk/hemelb}. \rev{Simulations were run 
either locally (Section \ref{sec:results}\ref{sec:simul_val}) or on 
up to 5040 cores of ARCHER, UK National Supercomputing Service 
(Section \ref{sec:results}\ref{sec:real_sim}, \ref{ap:grid_refine}, 
and \ref{ap:sens_analy})}

\begin{figure}
\centering
\includegraphics[scale=0.4]{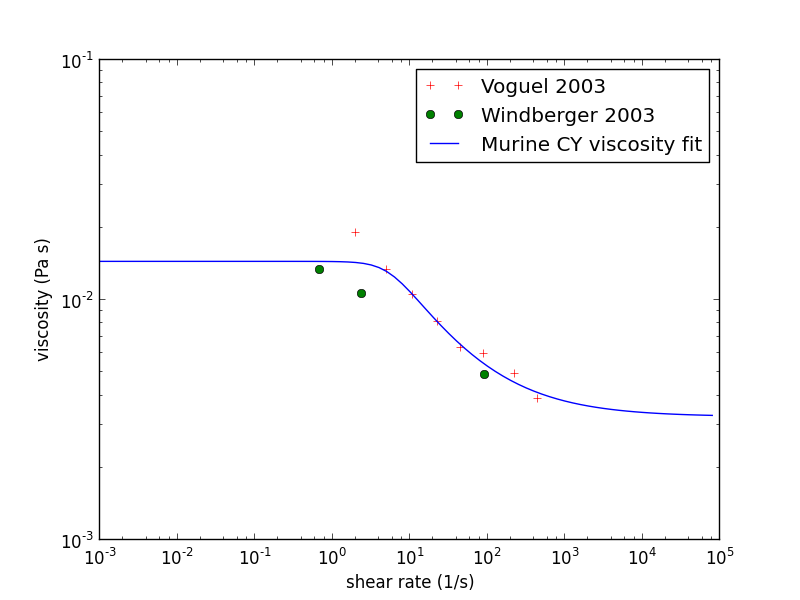}
\caption{Reported values of murine blood viscosity for different shear rates
  and Carreau-Yasuda (CY) model fit.
  \label{fi:cy_fit}}
\end{figure}

Experimental measurements were used to derive a functional form for
$\eta(\dot{\gamma})$ based on the Carreau-Yasuda (CY) mathematical
model (see \emph{e.g.} \citep{Boyd2007}):
\begin{equation}
\eta(\dot{\gamma}) = \eta_{\infty} + ( \eta_0 - \eta_{\infty})\left[ 1
  + (\lambda \dot{\gamma})^a \right]^{\frac{n-1}{a}}\;,
\label{eq:cy_model}
\end{equation}
where $a$, $n$, and $\lambda$ are empirically determined to fit a
curve between regions of constant $\eta_{\infty}$ and $\eta_0$. This
model defines three rheological regimes: a Newtonian region of
viscosity $\eta_0$ for low shear rate, followed by a shear-thinning
region where $\eta$ decreases with $\dot{\gamma}$; finally, a second
Newtonian region of viscosity $\eta_\infty$ is defined for high shear
rates. Equation \eqref{eq:cy_model} was fitted to the data in Table
\ref{ta:visc_values} with the least squares algorithm implemented in
the \rev{gnuplot graphing utility\footnote{Gnuplot 4.6.3, \url{ 
http://www.gnuplot.info}}} giving the following
results: \rev{$\eta_0 = \SI{14.49}{\milli\Pa\second}$, 
$\eta_\infty = \SI{3.265}{\milli\Pa\second}$, 
$\lambda= \SI{0.1839}{\second}$, $a=2.707$,
$n=0.4136$}. Figure \ref{fi:cy_fit} plots the data fit.

One of the many challenges when simulating blood flow in open domains
--- such as the subset of retinal vasculature that we 
present in Figure \ref{fig:model_recons} --- is the impact of the
choice of inlet/outlet boundary conditions on the simulated
haemodynamics. Ideally, one would use experimental measurements of
flow rate and/or pressure in order to close the system. We could not
obtain these data experimentally and relied on the
data surveyed in Section
\ref{sec:lit_rev}\ref{sec:ret_haem}. Pressures at the inlet and outlet
were set to the values measured by \citet{Wright2008} and presented in
Table \ref{ta:map_iop}: \SI{68.2}{\mmHg} and \SI{11.6}{\mmHg},
corresponding to the mean arterial pressure of the central retinal
artery and vein, respectively.

The LB algorithm admits a number of different implementations of the
no-slip boundary condition at the walls (see \emph{e.g.}
\citet{Latt2008, Nash2014} for surveys). We choose the method proposed by
\citet{Bouzidi2001} based on previous validation work
\citep{Nash2014}. In the current work, we perform further validation
as described in the following section. Finally, we initialise the
domain to a uniform density fluid at rest. The implications of this
choice are discussed in Section \ref{sec:results}\ref{sec:model_recons}.

\subsection{\rev{Code verification} methodology}
\label{sec:validation}

In this study, we are interested in using HemeLB to simulate blood
flow in a network of vessels of variable diameter, with differences of
up to one order of magnitude (see Table \ref{ta:vels_flows}).  This is
a challenging scenario since we must ensure that the spatial
discretisation is fine enough to capture all the features in the
capillaries and resolve flow accurately, while keeping the problem
computationally tractable due to the large number of fluid sites
arising from the discretisation of larger vessels. Furthermore, we are
particularly interested in generating accurate estimates of wall shear
stress in vessels that are typically not aligned with the Cartesian
grid, which can be challenging for regular grid based
methods. For example, \citet{Stahl2010} measured shear stress errors of up to 35\%
in the vicinity of the wall for non-lattice aligned channel flow with
the LB algorithm and the so-called bounce-back implementation of the
no-slip boundary condition \citep{Ziegler1993}.

In the current section parameters are denoted with a tilde when given
in lattice units and without when given in physical units. The LB
timestep $\Delta t$ is used as a conversion factor for time and the
voxel size $\Delta x$ for space such that \emph{e.g.} diameter $D =
\tilde D \Delta x$.
We also introduce the LB relaxation parameter (which controls the
viscosity in the lattice, for more details refer to \citet{Chen1998})
\begin{equation}
\label{eq:tau_nu}
\tilde \tau = \frac{1}{2} + \frac{\tilde \nu}{\tilde c_s^2}\;,
\end{equation}
where $\tilde \nu$ is the kinematic viscosity in lattice units
\begin{equation}
\label{eq:nu_lattice}
\tilde \nu=\frac{\eta}{\rho} \frac{\Delta t}{\Delta x^2}\;,
\end{equation}
and $\tilde c_s^2=\frac{1}{3}$ in the version of LB employed.

In the absence of experimental flow measurements to compare against
our computer simulations, we propose setting up a set of
benchmark simulations that capture the main flow and domain
characteristics and compare the results against known analytical
solutions. We will restrict ourselves to the simulation of steady,
Newtonian, and laminar flow in non-lattice aligned cylinders of
diameter $\tilde D\in[3,30]$ and length $\tilde L=4\tilde D$. The
orientation of the cylinder $\unitvec{n}_c$ is chosen pseudorandomly
from the unit sphere, subject to the constraint that $\unitvec{n}_c
\cdot \unitvec{e}_i \le 0.9$, $\forall i$. The value is
\begin{equation}
  \unitvec{n}_c = (-0.299, 0.382, 0.874)^\top\;.
\end{equation}
The laminar flow assumption is based on the Reynolds
numbers reported in the literature for microcirculation (\emph{e.g.}
$Re=0.2,0.05, \text{and } 0.0003$ for arterioles, venules, and
capillaries, respectively \citep{Popel2005}). We choose $Re=1$ in our
validation. The steady and Newtonian flow assumptions are made to
simplify the analytical solution of the benchmarks considered. Their
implications are discussed in Section \ref{sec:results}\ref{sec:simul_val}.
Finally, a parabolic velocity profile with maximum
velocity
\begin{equation}
\vec{\tilde v}_{\text{max}} = -\frac{\tilde \nu Re}{\tilde D} \unitvec{n}_c \;,
\end{equation}
is imposed at the inlet \citep{Ladd1994}.

The purpose of our validation study is two-fold. First, to
characterise the accuracy of the recovered haemodynamics as a function
of the number of fluid sites across a given vessel. Second, to
evaluate the accuracy of the computed wall shear stress given our
choice of implementation of the no-slip boundary condition
\citep{Bouzidi2001}, which has --- to the best of our knowledge --- not
been done before.

In our first experiment, we compare the volumetric flow rate
\begin{equation}
q = \iint_S \vec{v}\cdot \unitvec{n}\, \mathrm{d}S
\end{equation}
integrated over a lattice aligned cross-sectional plane (defined by
point $(0,0,0)^\top$ and plane normal $(0,0,1)^\top$) with
the analytical solution of Hagen-Poiseuille flow in an infinite
cylinder
\begin{equation}
q^{\star}=\frac{|\vec{v}_{\text{max}}|\pi D^2}{8}\;,
\end{equation}
for a range of values of diameter $\tilde D\in[3,30]$ and relaxation time
$\tilde \tau\in\{0.6,0.8,1,1.2,1.4\}$.

In the second experiment, we compare $\mat{T}$ with
the --- appropriately rotated --- analytical solution of the
Hagen-Poiseuille shear stress tensor $\mat{T}'$ in a cylinder of axis
$\unitvec{e}_3=(0,0,1)^\top$ and radius $R=D/2$ assuming flow in the \rev{positive} direction
of the cylinder axis. For a given point in the domain $\vec{x}=(x_1,
x_2, x_3)^\top$ such that $x_1^2+x_2^2 = r^2$, $r \in [0,R]$, it can be
shown that
\begin{equation}
  \mat{T}'(\vec{x}) = 
  % \left(
  %   \begin{tabular}{ccc}
  %     0 & 0 & -$\frac{\Delta P x_1}{2L}$ \\
  %     0 & 0 & -$\frac{\Delta P x_2}{2L}$ \\
  %     -$\frac{\Delta P x_1}{2L}$ & -$\frac{\Delta P x_2}{2L}$ & 0
  %   \end{tabular}
  % \right) = 
  \left(
    \begin{tabular}{ccc}
      0 & 0 &  -$2|\vec{v}_{\text{max}}|\eta R^{-2}x_1$ \\
      0 & 0 & -$2|\vec{v}_{\text{max}}|\eta R^{-2}x_2$ \\
      -$2|\vec{v}_{\text{max}}|\eta R^{-2}x_1$ & -$2|\vec{v}_{\text{max}}|\eta R^{-2}x_2$ & 0
    \end{tabular}\right)\;.
\end{equation}
% where $\Delta P$ is the pressure difference between the cylinder inlet
% and the outlet given by
% \begin{equation}
% \frac{\Delta P}{L} = \frac{4|\vec{v}_{\text{max}}|\eta}{R^2}\;.
% \end{equation}
The tensor rotation is defined by the matrix $\mat{R}=(\unitvec{r}_1\ \unitvec{r}_2\ \unitvec{r}_3)$,
\begin{align}
\unitvec{r}_1 &= \unitvec{n}_c, \\
\unitvec{r}_2 &= \frac{\unitvec{n}_c \times \unitvec{e}_3}{||\unitvec{n}_c \times \unitvec{e}_3||} , \\
\unitvec{r}_3 &= \unitvec{r}_1 \times \unitvec{r}_2,
\end{align}
where $||\cdot||$ is the magnitude of a vector, such that
\begin{equation}
\mat{T}^{\star} = \mat{R}\mat{T}'\mat{R}^\top\;.
\end{equation}

\section{\rev{Results and discussion}}
\label{sec:results}

\subsection{\rev{Code verification}}
\label{sec:simul_val}

\begin{figure}
\centering
\includegraphics[scale=0.5]{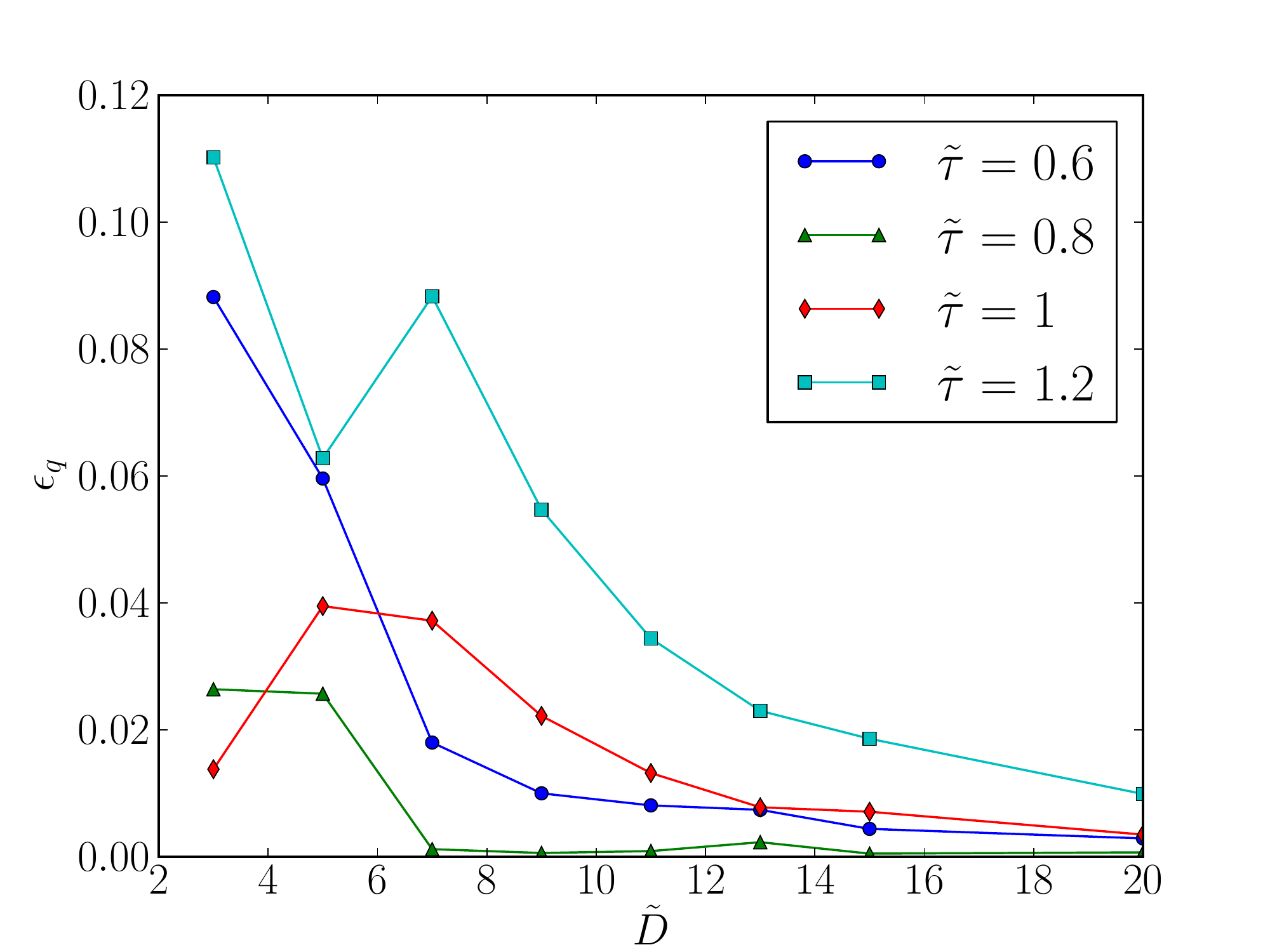}
\caption{Hagen-Poiseuille flow in an inclined cylinder. Relative error
  on the computed flow rate as a function of vessel diameter $\tilde
  D$ and lattice-Boltzmann (LB) relaxation time $\tilde \tau$. For $\tilde \tau=0.8$
  the total error is kept below 3\% even for cylinders with just 3
  lattice sites across. These results confirm the suitability of the
  LB algorithm for the simulation of flow in sparse geometries and
  porous media. The lines are a guide to the eye and bear no physical
  meaning.}
\label{fig:flow_rate}
\end{figure}

Figure \ref{fig:flow_rate} plots the relative error in the simulated
flow rate
\begin{equation}
\epsilon_q = \left\lvert\frac{q^{\star}-q}{q^{\star}} \right\lvert
\end{equation}
as a function of the cylinder diameter.
We observe how the choice of $\tilde \tau$ greatly affects the accuracy of
the simulated haemodynamics. In agreement with similar analyses in the
literature (see \emph{e.g.} \citet{Kruger2009}), the error is larger
for values close to the stability threshold of $\tilde \tau=0.5$ and for
values greater than 1. A region of excellent accuracy is located
around $\tilde \tau=0.8$. In that case, the relative error $\epsilon_q$ stays
below 3\% for all the values of $\tilde D$ studied. The ability to
simulate correct flow dynamics in channels with only a few lattice
sites across has gained the LB algorithm wide acceptance for the
simulation of flow in complex domains and porous media
\citep{Aidun2010}.

\begin{figure}
  \centering 
  \subfigure[$\tilde D=5$, analytical vs. computed solution.]{
    \label{fig:shear_stress:d5_anal_vs_comp}
    \includegraphics[scale=0.25]{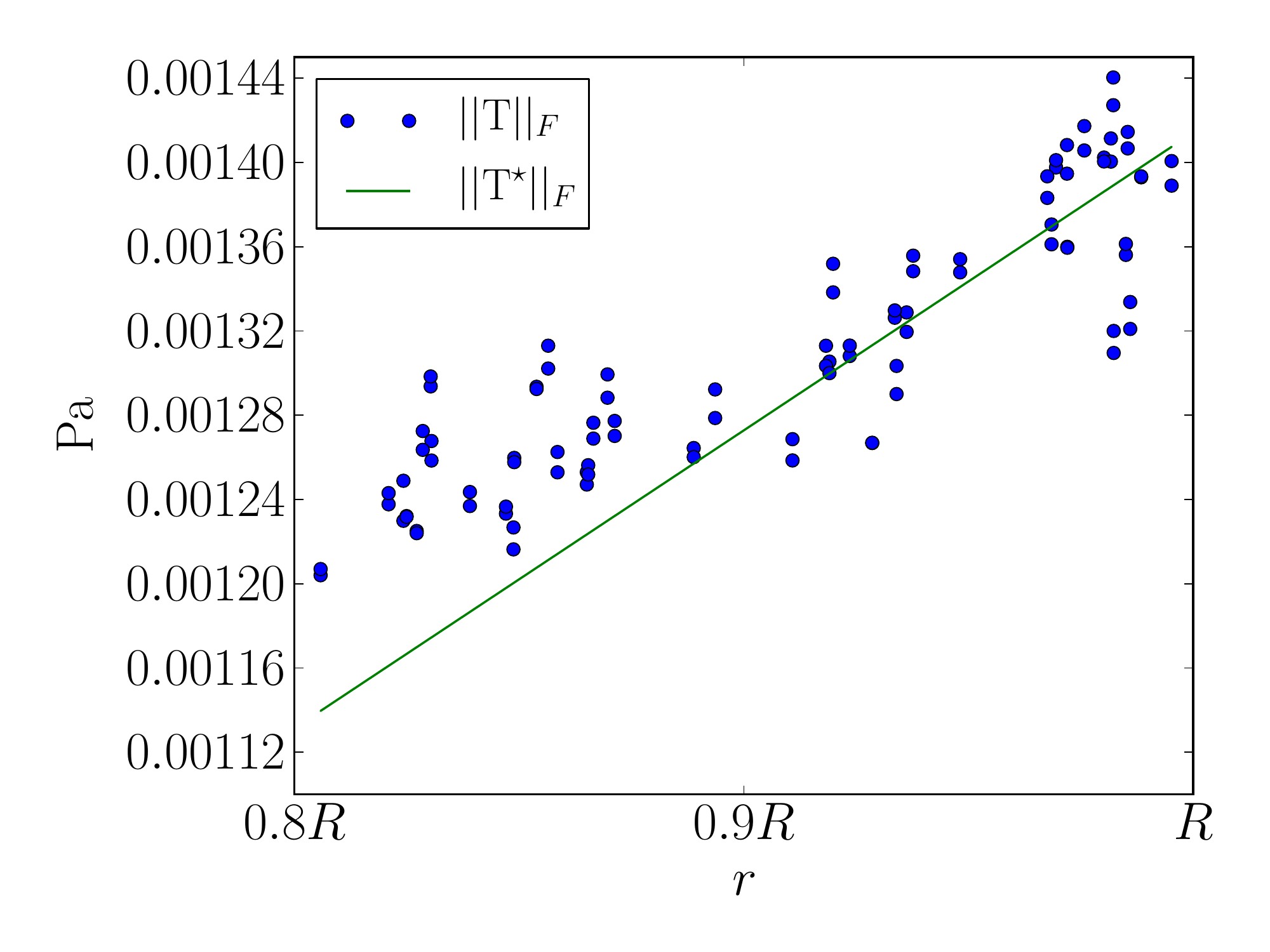}
  }
  \subfigure[$\tilde D=5$, relative error.]{
    \label{fig:shear_stress:d5_err}
    \includegraphics[scale=0.25]{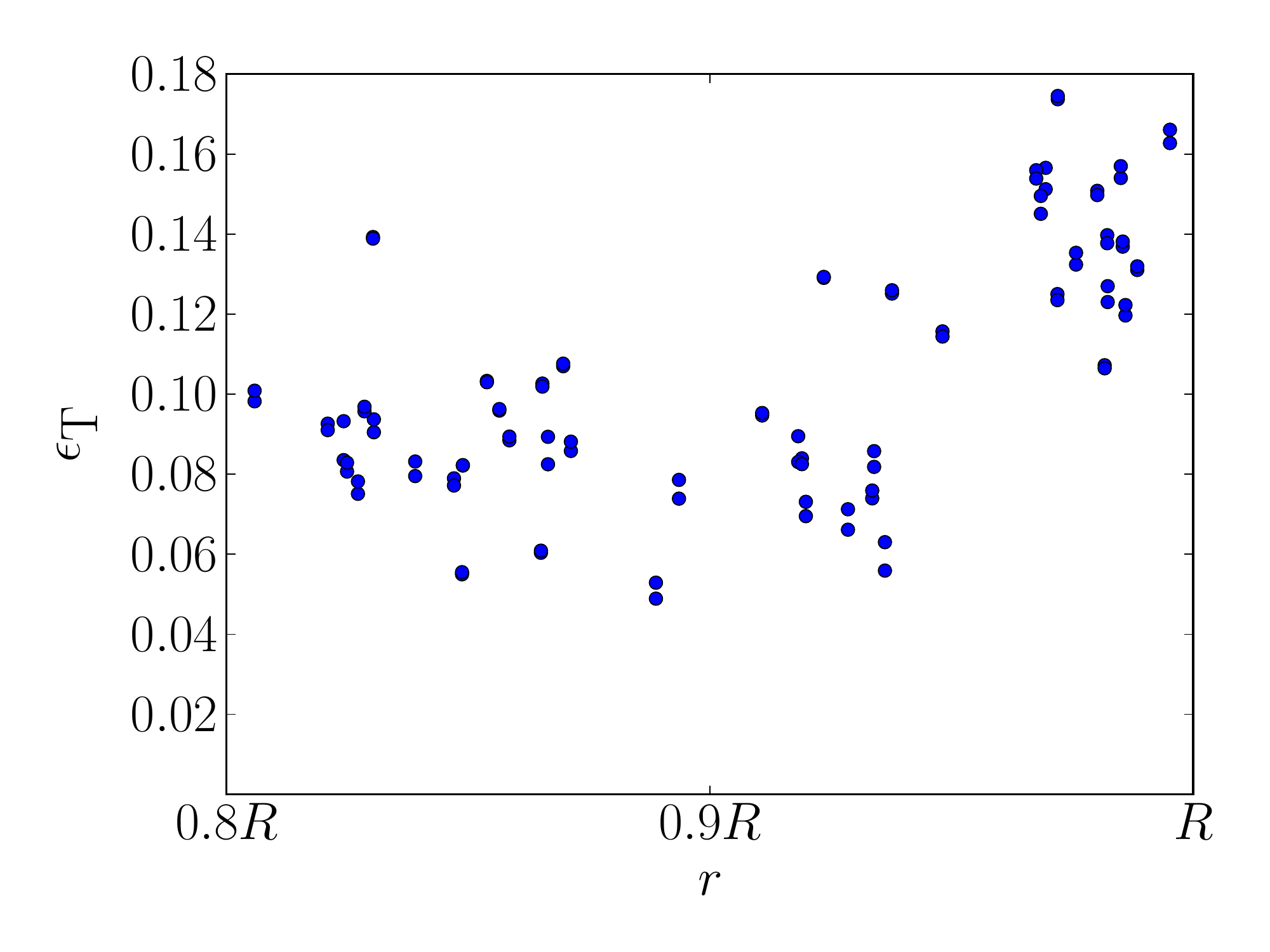}
  }
  \subfigure[$\tilde D=7$, analytical vs. computed solution]{
    \label{fig:shear_stress:d7_anal_vs_comp}
    \includegraphics[scale=0.25]{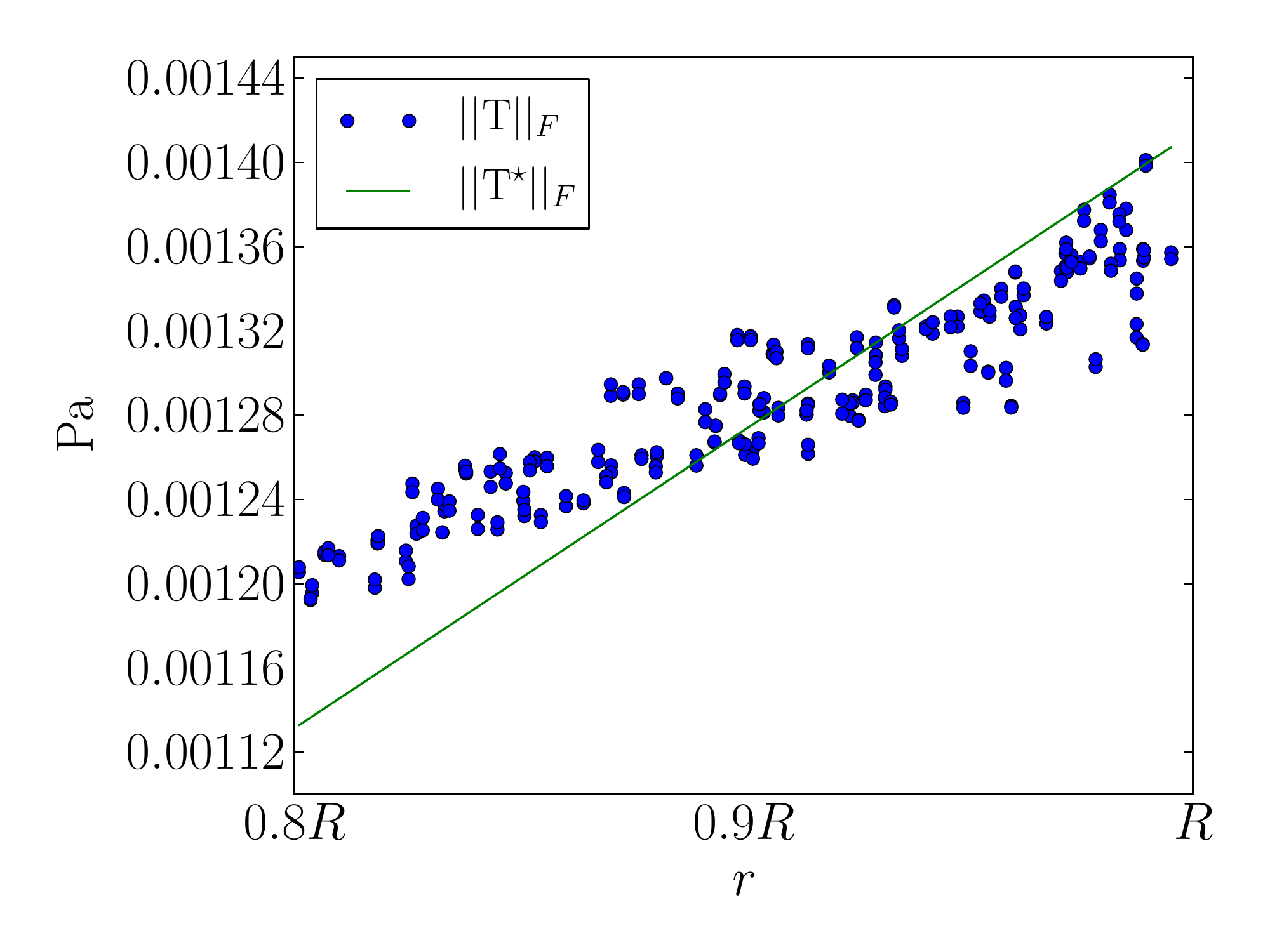}
  }
  \subfigure[$\tilde D=7$, relative error.]{
    \label{fig:shear_stress:d7_err}
    \includegraphics[scale=0.25]{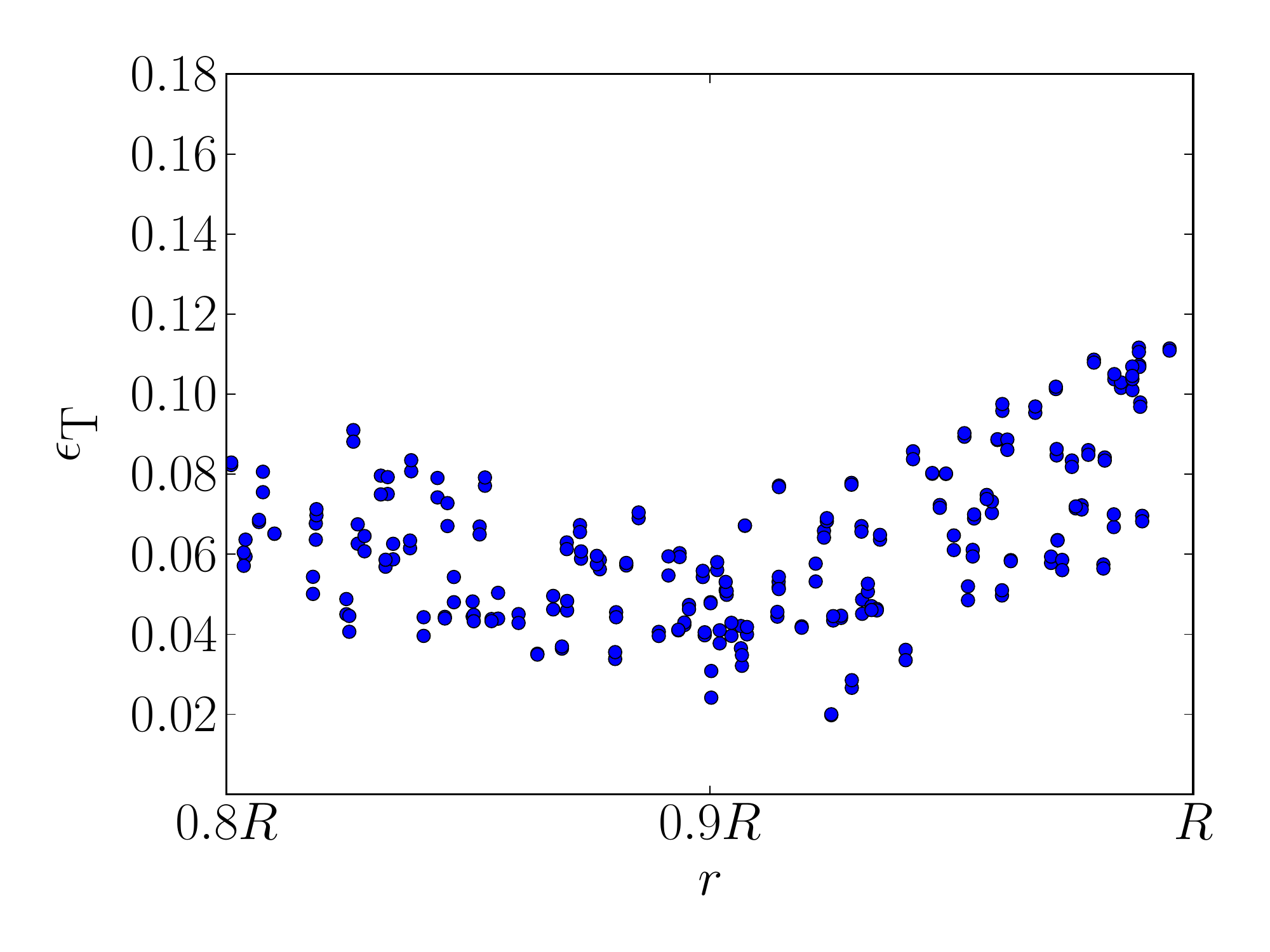}
  }
%   \subfigure[$\tilde D=10$, analytical vs. computed solution]{
%     \label{fig:shear_stress:d10_anal_vs_comp}
% %    \includegraphics[scale=0.25]{results/d10_comp_vs_ana_final}
%     \includegraphics[scale=0.25]{d10_comp_vs_ana_final}
%   }
%   \subfigure[$\tilde D=10$, relative error.]{
%     \label{fig:shear_stress:d10_err}
% %    \includegraphics[scale=0.25]{results/d10_rel_error_final}
%     \includegraphics[scale=0.25]{d10_rel_error_final}
%   }
  \subfigure[$\tilde D=15$, analytical vs. computed solution.]{
    \label{fig:shear_stress:d15_anal_vs_comp}
    \includegraphics[scale=0.25]{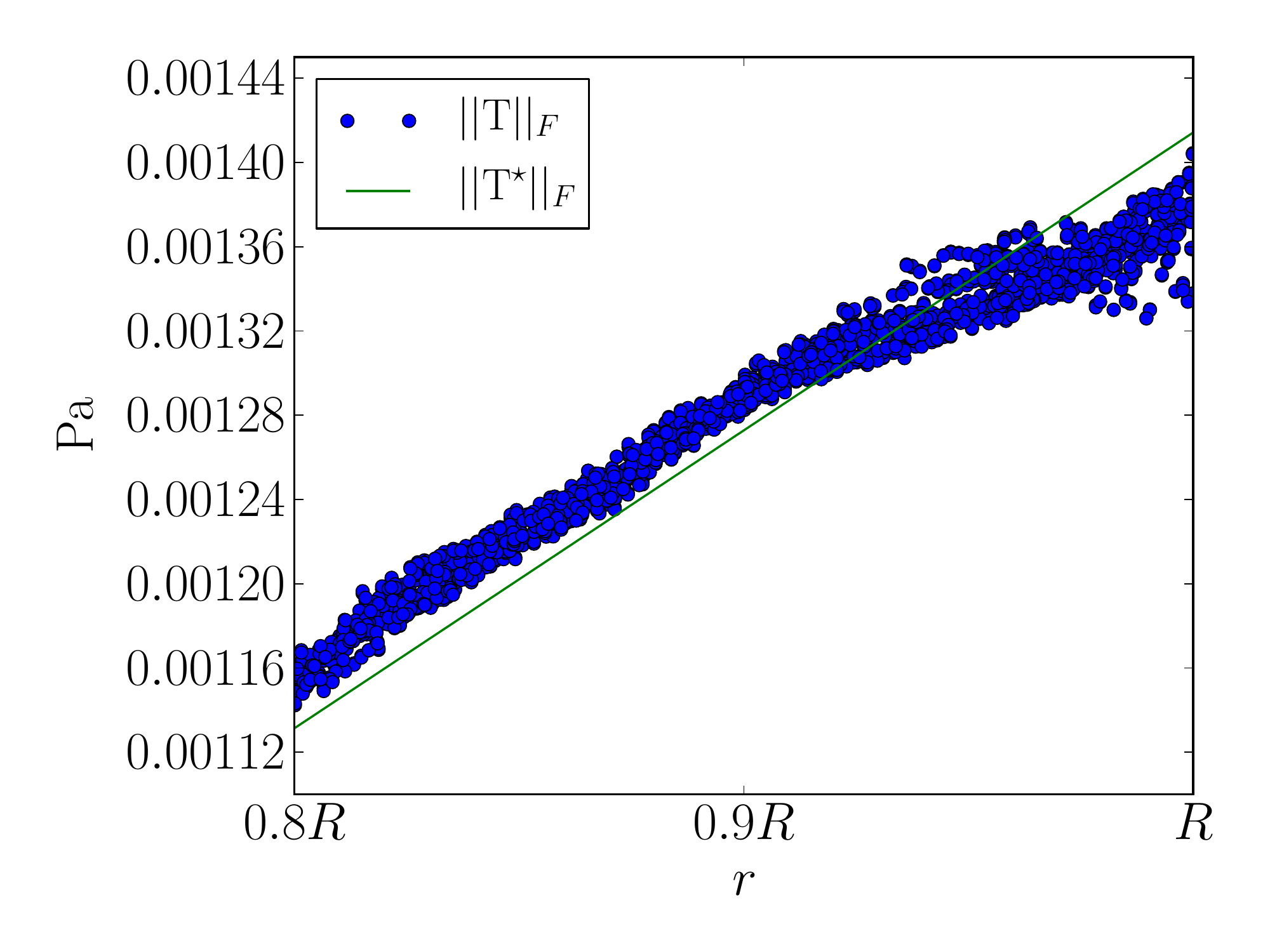}
  }
  \subfigure[$\tilde D=15$, relative error.]{
    \label{fig:shear_stress:d15_err}
    \includegraphics[scale=0.25]{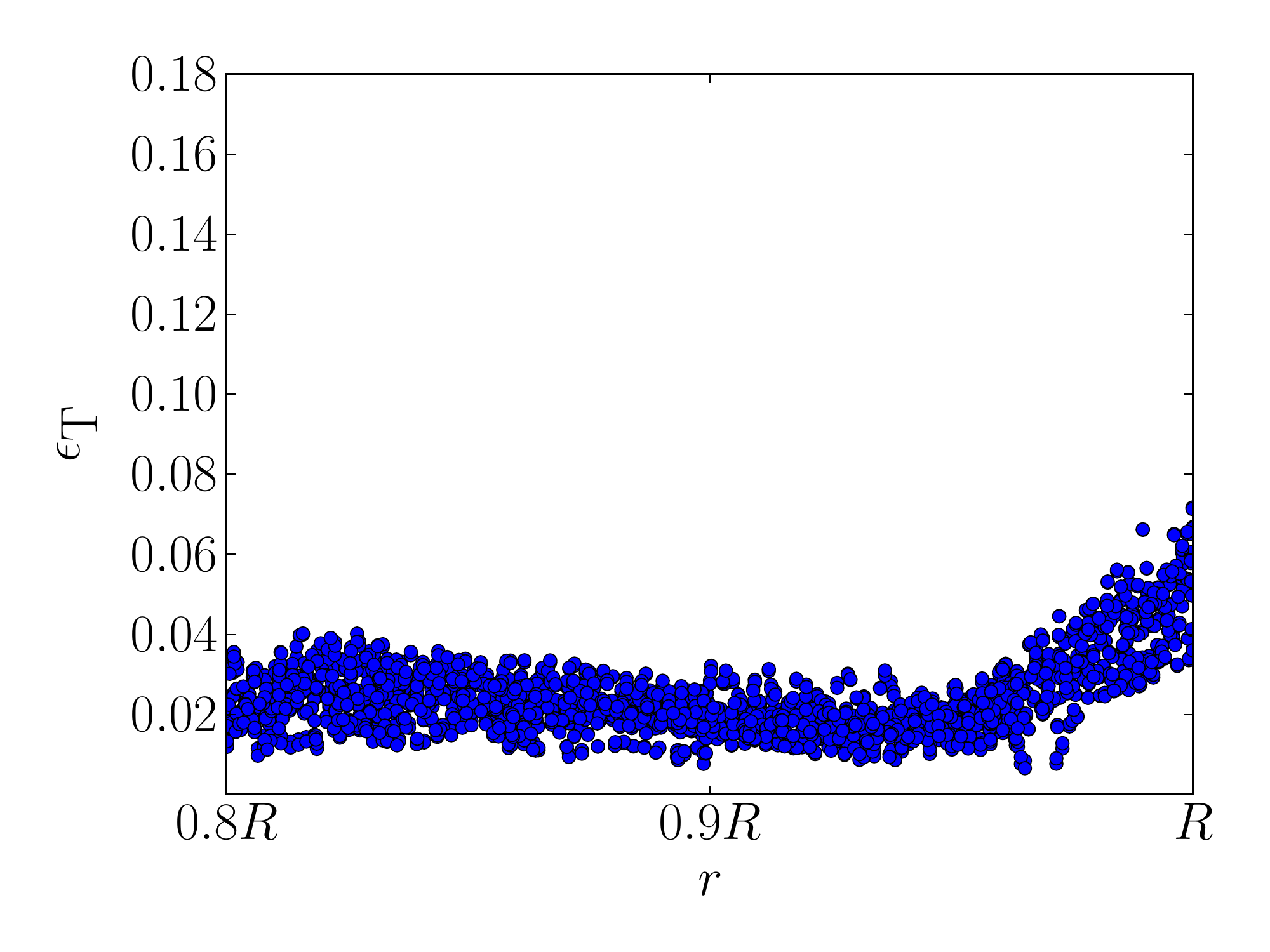}
  }
  \caption{Hagen-Poiseuille shear stress in inclined cylinders of
    $\tilde D\in \{5,7,15\}$. Norm of the analytical and computed
    stress tensors (panels \ref{fig:shear_stress:d5_anal_vs_comp},
    \ref{fig:shear_stress:d7_anal_vs_comp},
    \ref{fig:shear_stress:d15_anal_vs_comp}) and relative error
    between them (panels \ref{fig:shear_stress:d5_err},
    \ref{fig:shear_stress:d7_err},
    \ref{fig:shear_stress:d15_err}). Results are presented for every
    lattice site with radius $r\in [0.8R, R]$. Agreement between
    computed and analytical solution improves with increasing $\tilde
    D$ and it is always better in the vicinity of the cylinder
    wall. These results, with the \citet{Bouzidi2001} implementation
    of the no-slip boundary condition, represent a substantial
    improvement over the 35\% error reported by \citet{Stahl2010} with
    the bounce-back method and $\tilde D=20$.}
\label{fig:shear_stress}
\end{figure}

Figure \ref{fig:shear_stress} plots, for a range of values of $\tilde
D$, the computed and analytical solutions of the shear stress tensor
($\mat{T}$ and $\mat{T}^{\star}$) as well as the associated relative error
\begin{equation}
\label{eq:T_rel_error}
\epsilon_\mat{T} = \frac{||\mat{T}^{\star}-\mat{T}||_F}{||\mat{T}^{\star}||_F}\;,
\end{equation}
where $||\cdot||_F$ is the Frobenius norm of an $m\times n$ matrix
$\mat{A}=[a_{ij}]$
\begin{equation}
||A||_F = \sqrt{\displaystyle \sum_{i=1}^m \sum_{j=1}^n |a_{ij}|^2}\;.
\end{equation}
The choice of error norm in \eqref{eq:T_rel_error} ensures that error
contributions given by individual components cannot compensate, hence
being sensitive to rotation errors. We are mainly interested in the
accuracy of the shear stress calculation in the vicinity of the vessel
wall. Therefore, the results are presented for the subdomain defined
by all the lattice sites with $r \in [0.8R, R]$.

It can be observed how the recovered shear stress follows the expected
pattern of monotonic increase from zero at the cylinder axis (results
not shown here) to its maximum value at the wall. However, a certain
deviation exists compared to the analytical solution. For values of
$\tilde D\geq 7$, we observe a clear pattern of shear stress being
overestimated in the $[0.8R,0.9R]$ region while being underestimated
in $[0.9R,R]$. Panels \ref{fig:shear_stress:d5_err},
\ref{fig:shear_stress:d7_err}, and
\ref{fig:shear_stress:d15_err} quantify this error. We observe how the
largest error in the domain always occurs at the cylinder wall and that
it decreases as $\tilde D$ increases: from around 17\% for $\tilde D=5$ to around 7\%
for $\tilde D=15$. These results, with the \citet{Bouzidi2001} implementation
of the no-slip boundary condition, represent a substantial improvement
over the 35\% error reported by \citet{Stahl2010} with the bounce-back
method and $\tilde D=20$, confirming the superiority of the former algorithm. More
importantly, the wall shear stress is consistently underestimated and
the error decreases for $\tilde D\geq 7$. Therefore, one could implement an
\emph{a posteriori} correction of the computed shear stress tensor
based on this knowledge. Such a development is beyond the scope of the current
paper and will be developed as part of a future study.

It is interesting to note that the errors in flow rate and shear stress are,
to a large extent, decoupled. This is due to the characteristics of the LB
algorithm as detailed in \ref{ap:lb}. In our work we will define two rules for
the accurate simulation of blood flow in our network of vessels of variable
diameter. First, a minimum diameter of $\tilde D=3$ will be enforced
throughout the domain. This will ensure that, for $\tilde \tau=0.8$, the
general flow patterns produced are accurate. Second, values of shear stress in
regions of interest will only be considered valid if $\tilde D\geq 7$. The
error estimates reported in Figure \ref{fig:flow_rate} will be taken into
account in the presentation of our results.

Finally, we turn our attention to the implications of the use of a
generalised Newtonian rheology model in our retinal flow simulations.
In the current section we observed that $\tilde \tau=0.8$ minimises the error
in the flow rate recovered. In the case of Newtonian fluids,
regardless of the value of viscosity being simulated, one can choose
$\frac{\Delta t}{\Delta x^2}$ such that Equation \eqref{eq:tau_nu}
yields the desired value of $\tilde \tau$ (this will obviously have an impact
on computational cost). For generalised Newtonian fluids, $\tilde \tau$
becomes a function of $\dot{\gamma}$ and will take values in the range
$[\tilde \tau_\infty, \tilde \tau_0]$. In the rheology model presented in Section
\ref{sec:methods}\ref{sec:sim_setup}, $\frac{\nu_0}{\nu_\infty}=4.92$,
which would lead to impractical values of $\tilde \tau_\infty$ unless
$\tilde \tau_0$ is chosen small enough. We will therefore choose the
coefficient $\frac{\Delta t}{\Delta x^2}$ such that $\tilde \tau_\infty=0.6$
and $\tilde \tau_0=0.992$. Based on the
results in Figure \ref{fig:flow_rate}, this choice will yield an error
of less than 4\% in the flow rate recovered for $\tilde D\geq 7$.

\subsection{Model reconstruction}
\label{sec:model_recons}

\begin{figure}
  \centering 
  \subfigure[]{
    \label{fig:retina_img}
    \includegraphics[scale=0.0170]{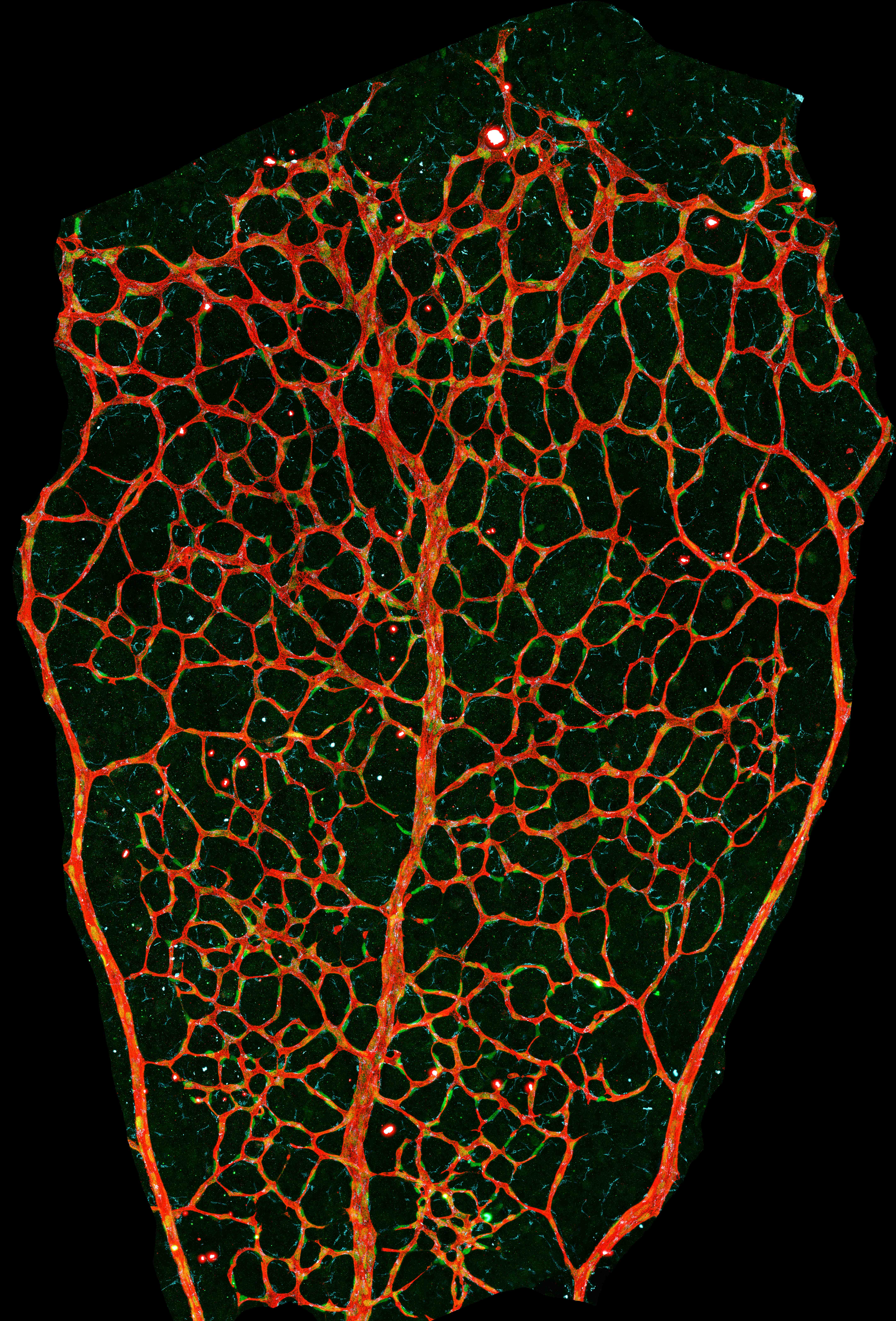}
  }
  \subfigure[]{
    \label{fig:segmen}
    \includegraphics[scale=0.25]{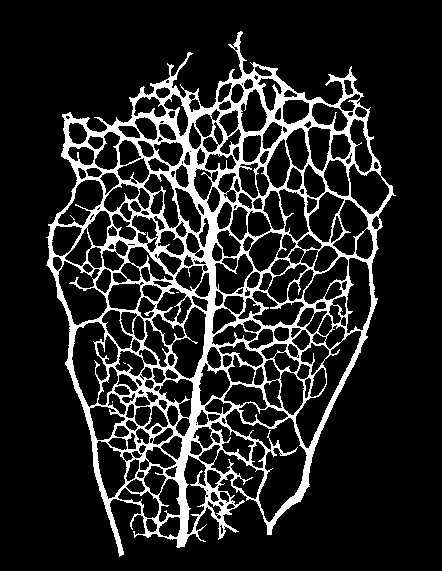}
  }
  \subfigure[]{
    \label{fig:recons}
    \includegraphics[scale=0.07]{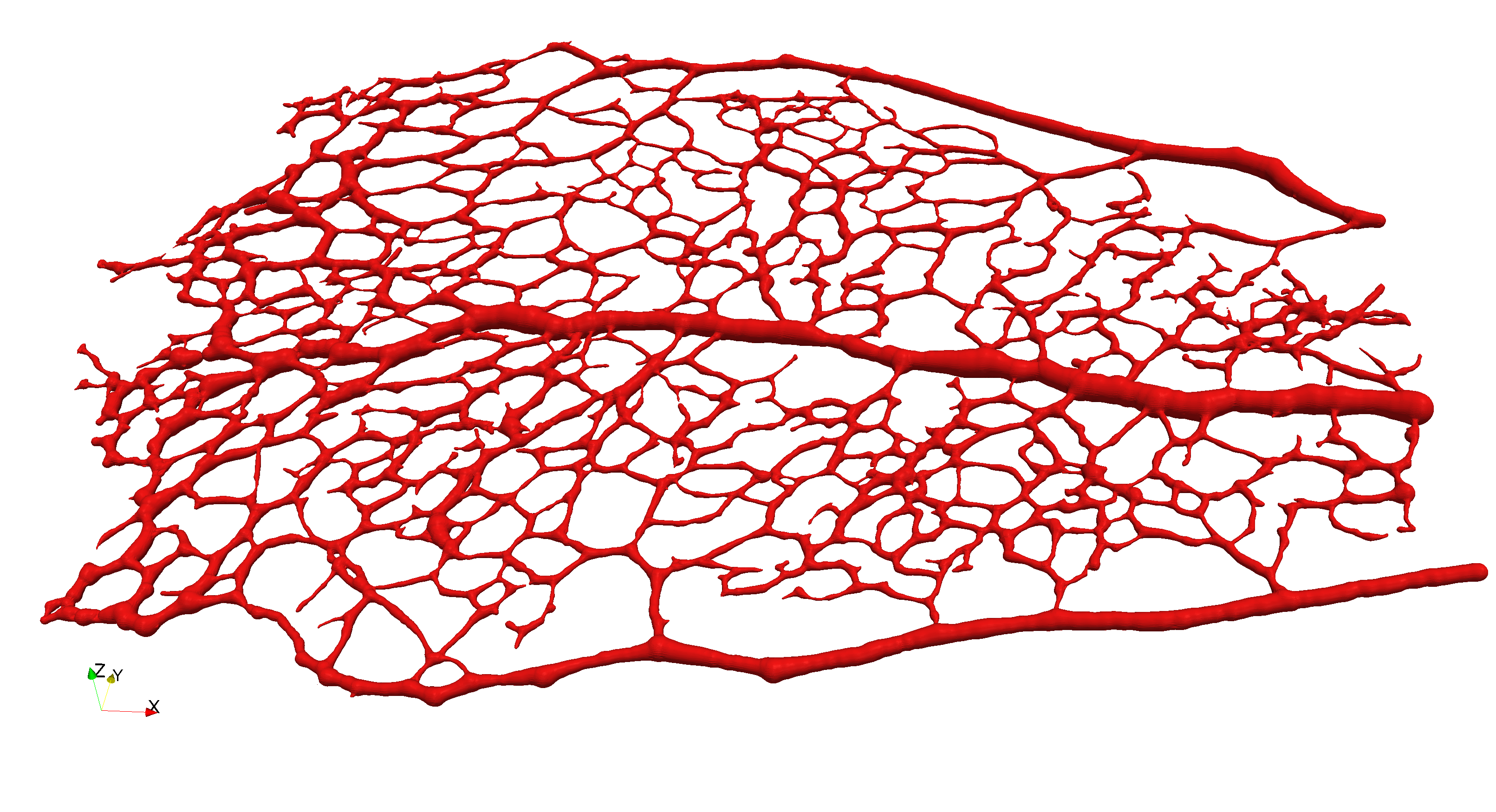}
  }
  \caption{Subset of a wildtype P6 retinal plexus used to reconstruct
    one of our retinal blood flow models, \rev{namely P6A model}. The
    original microscope image is segmented and the network skeleton
    and segment radii are computed. Based on these values, a 3D volume
    is reconstructed assuming vessels of piece-wise constant
    radius. (a) Original image. (b) Segmented image. (c) Reconstructed
    surface.}
\label{fig:model_recons}
\end{figure}

\begin{figure}
  \centering 
  \subfigure[P5A model.]{
    \label{fig:p5_1}
    \includegraphics[scale=0.08]{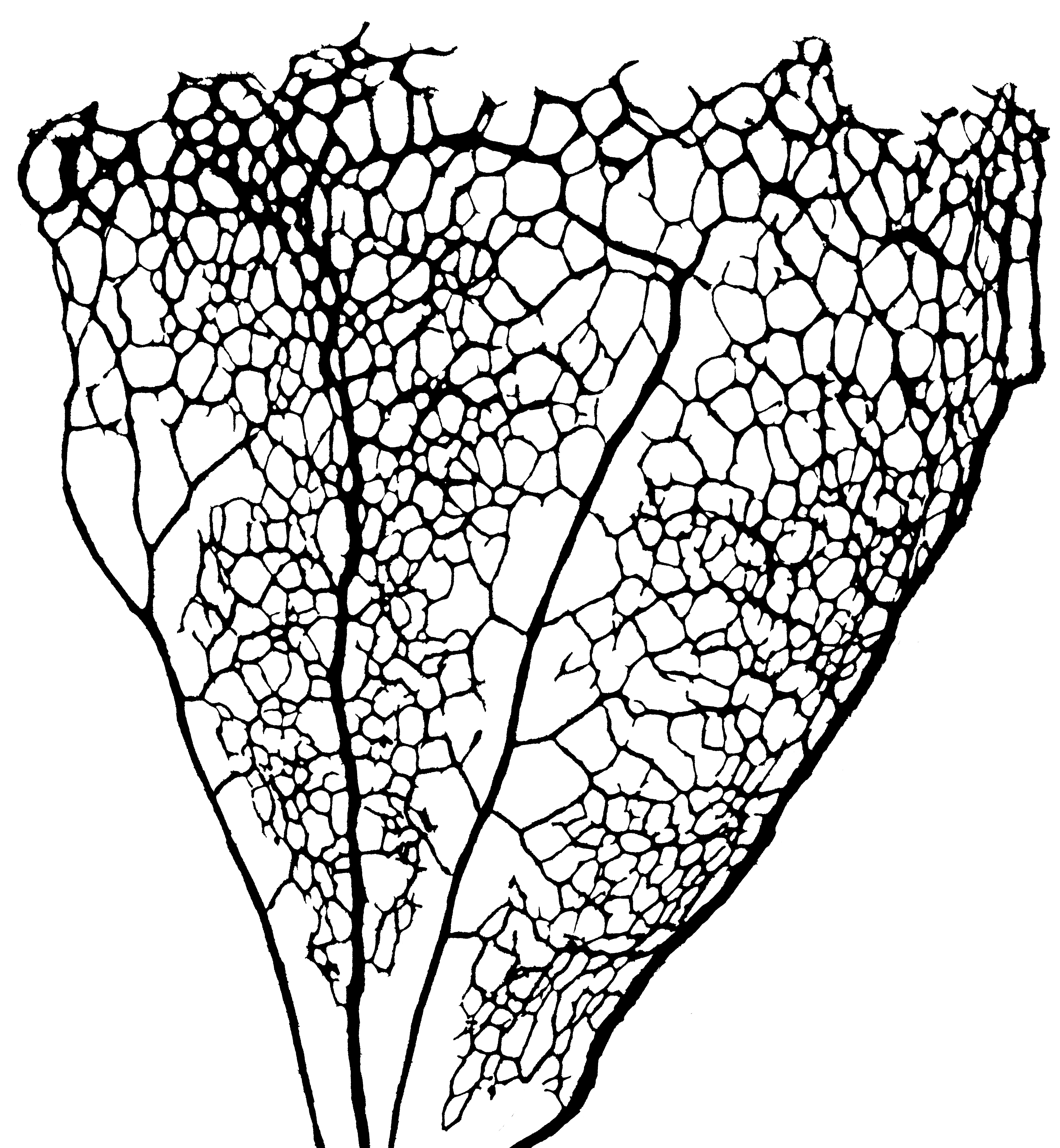}
  }
  \subfigure[P5B model.]{
    \label{fig:p5_2}
    \includegraphics[scale=0.02]{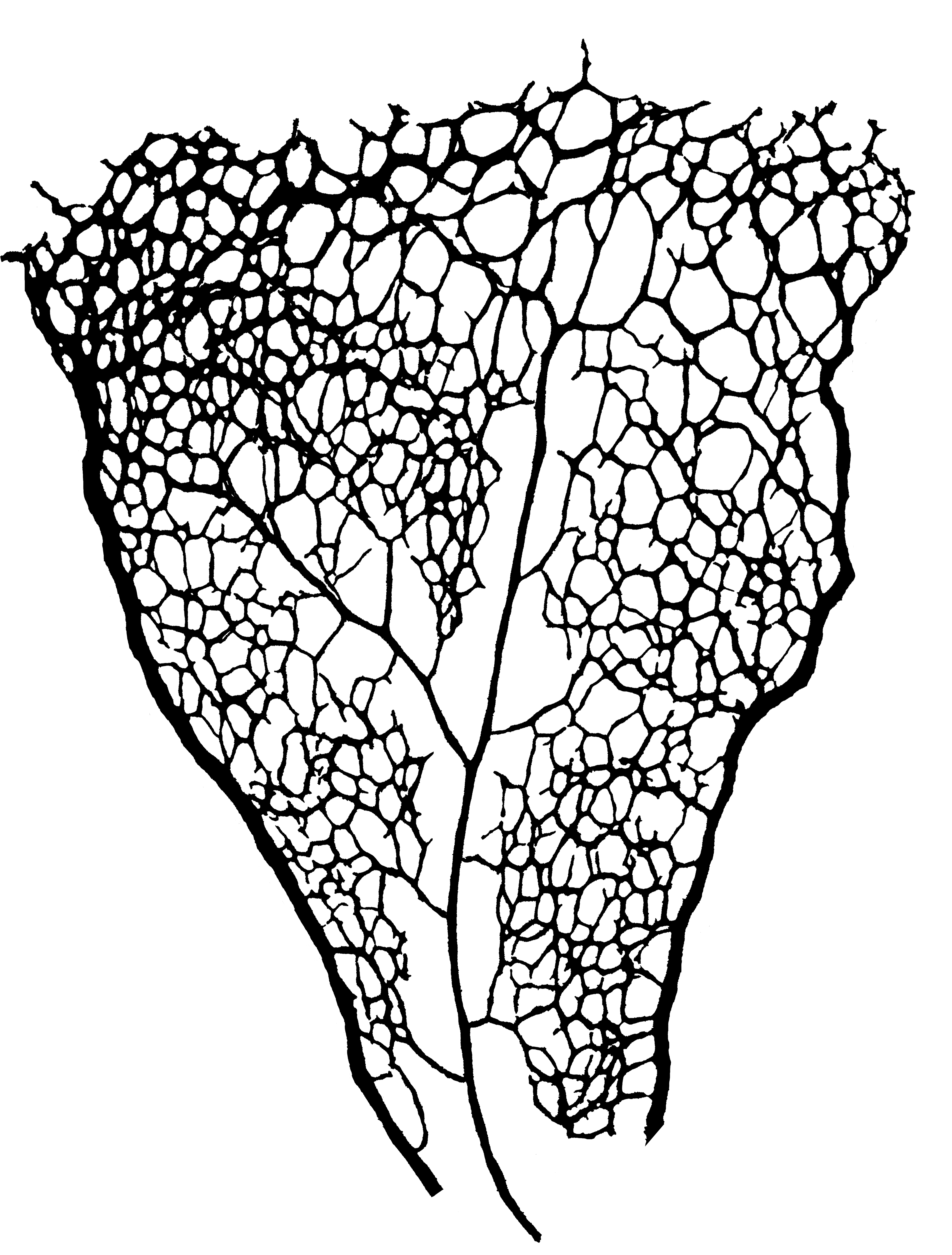}
  }
  \subfigure[P6B model.]{
    \label{fig:p6_1}
    \includegraphics[scale=0.015]{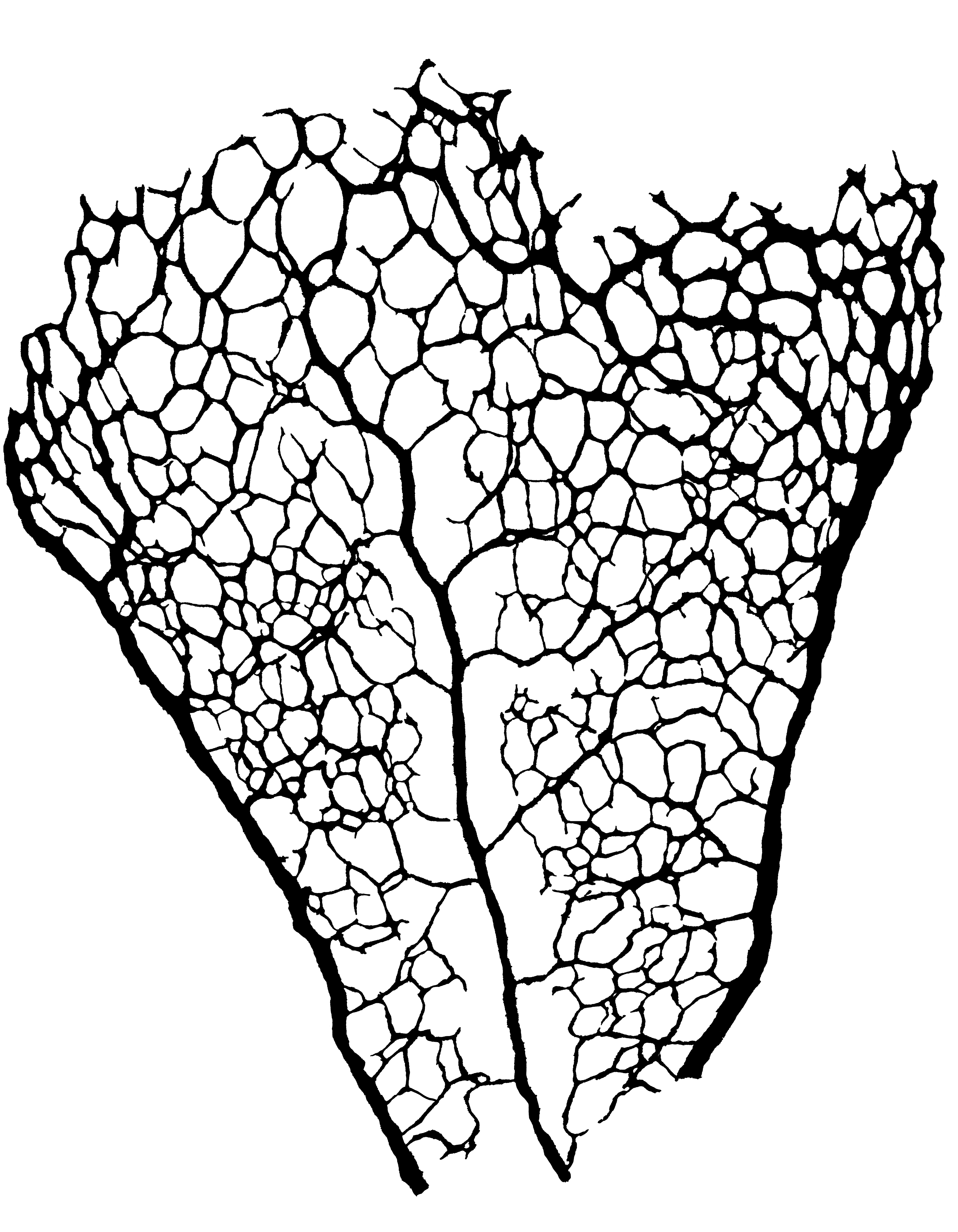}
  }
  % \subfigure[P6 wildtype retina.]{
  %   \label{fig:p6_2}
  %   \includegraphics[scale=0.02]{WT-P6-2_5_312_segmented.png}
  % }
  \caption{\rev{Binary masks defining the luminal surface of three
      retinal plexuses obtained at two different stages of
      development. All plexuses are presented with the area closer to
      the optic disc at the bottom of the image and the sprouting
      front at the top. In all samples studied, arteries tend to be
      thinner and have less daughter vessels than veins. Vessels close
      to the sprouting front tend to have less well-defined identity
      with luminal diameters comparable to arteries/veins. This is
      particularly noticeable in the P5 samples. Vessel density is
      also higher close to the sprouting front in P5 retinas.}}
  \label{fig:more_retinas}
\end{figure}

The vascular plexus of wildtype retinas was stained with the luminal
membrane marker ICAM2 and
images were acquired using a confocal microscope as described
in Section \ref{sec:methods}\ref{sec:image_proc}.  Figure
\ref{fig:retina_img} shows a region of interest in one of the imaged
retinas. It contains, on either side, two arterial segments coming
from the optic disc and connecting with a segment of a retinal vein
(centre of the image) through a dense capillary network. It can be
appreciated how the network is more mature (\emph{e.g.}  vessel identity, branching patterns) closer to the optic disc
(bottom of the image) while its structure
is much more primitive and less remodelled in the periphery closer to
the sprouting front (top of the image). Figure \ref{fig:segmen} presents the results of
the image segmentation process. The algorithm described in Section
\ref{sec:methods}\ref{sec:image_proc} is used to first create a binary
mask separating the luminal area and background tissue and second
extract the network skeleton and radii. The latter are used to
reconstruct the 3D luminal surface under the assumption of vessel circular
cross-section (see Section \ref{sec:methods}\ref{sec:image_proc} for a
discussion). Figure \ref{fig:recons} shows the
reconstructed surface. \rev{We refer to this model as P6A.}
\rev{Figure \ref{fig:more_retinas} presents luminal surface binary
  masks for three additional P5 and P6 retinal plexuses. The same
  reconstruction algorithm is applied and the resulting models are
  referred to as P5A, P5B, and P6B, respectively.}

\begin{figure}
  \centering
  \subfigure[P5B model.]{
    \label{fig:hist_p5}
    \includegraphics[scale=0.45]{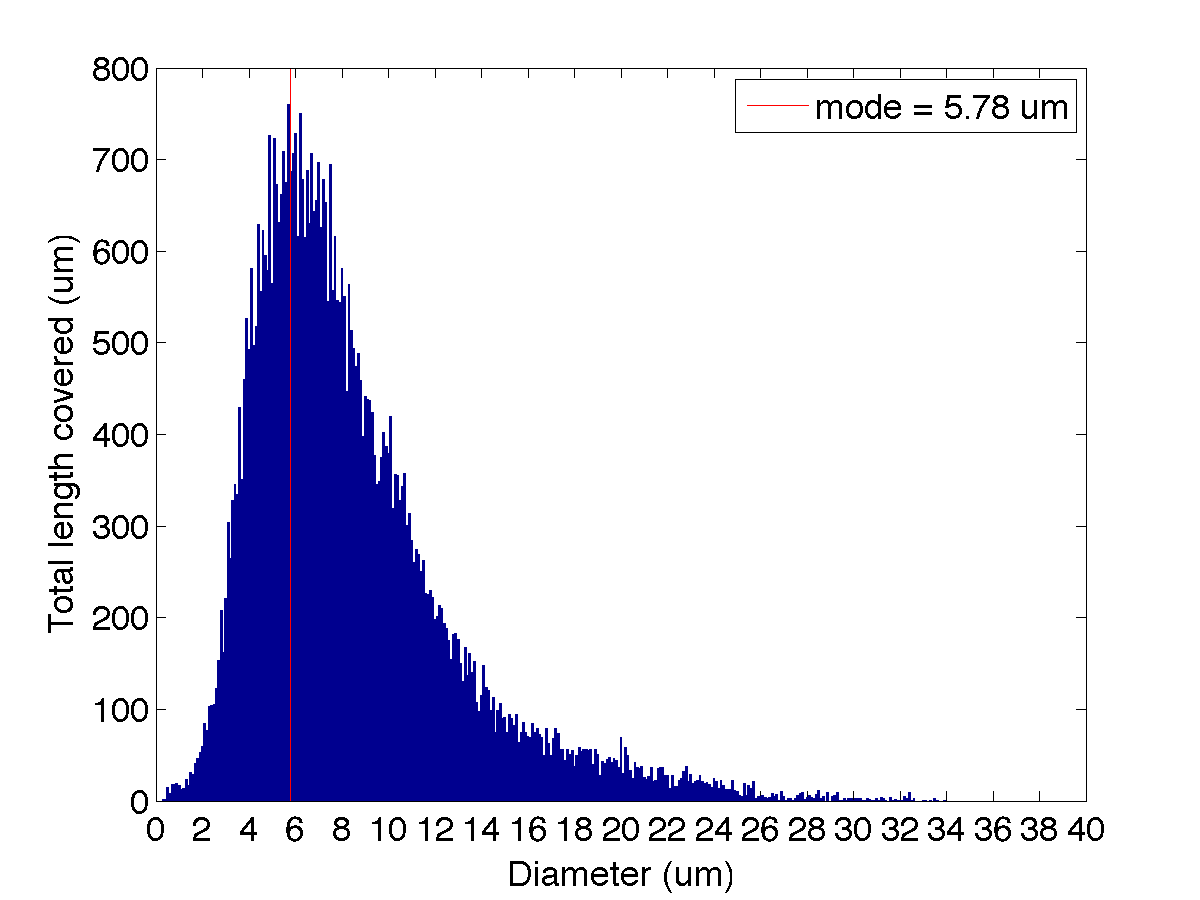}
  }
  \subfigure[P6B model.]{
    \label{fig:hist_p6}
  \includegraphics[scale=0.45]{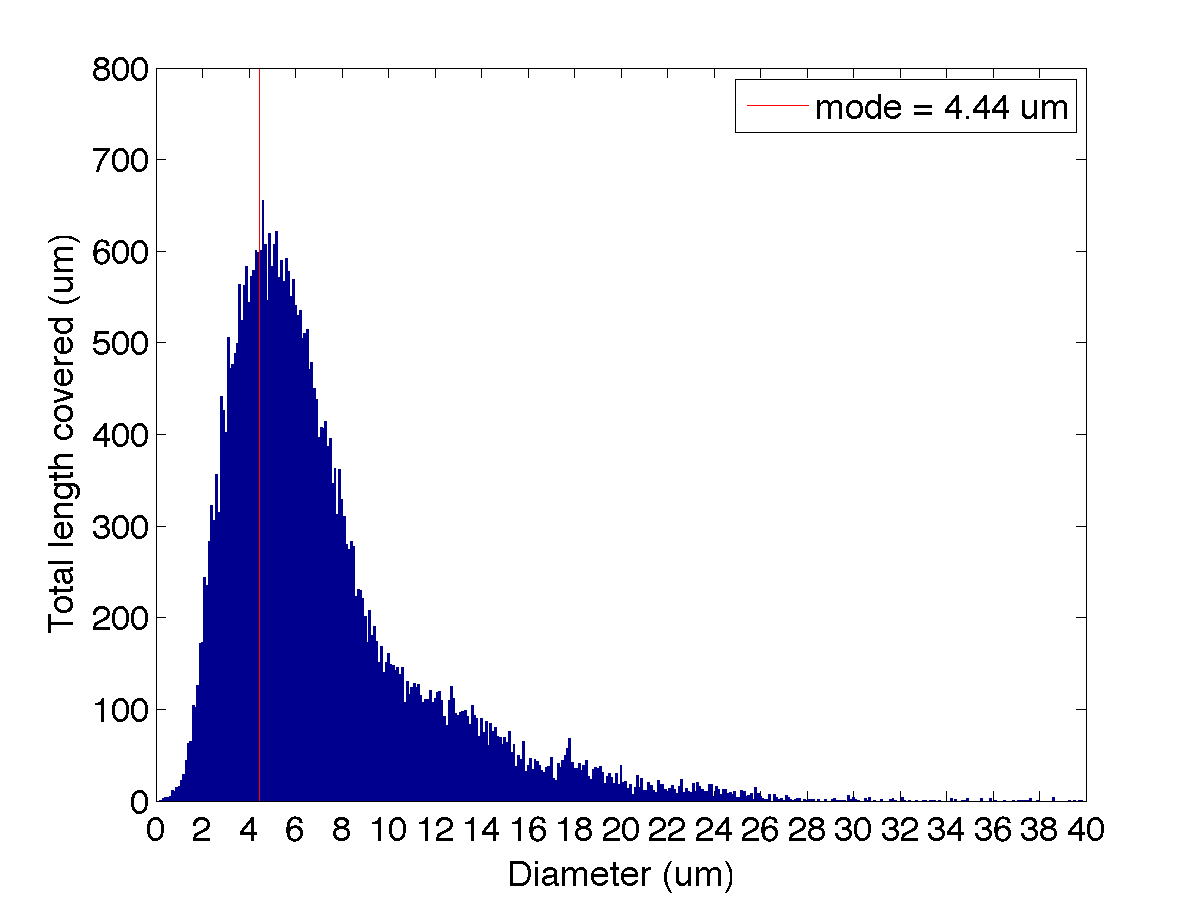}  
  }
  \caption{Network diameter histogram showing the aggregated total
    distance covered by vessels of a given diameter. \rev{Vertical 
    lines indicate the mode of a lognormal probability distribution 
    fit of each dataset. The values for the models not shown here are 
    \SI{5.51}{\micro\meter} (P5A) and \SI{5.29}{\micro\meter} (P6A). 
    We use these values as an estimate of the typical capillary diameter
    (the most common type of vessel in the network). Capillaries with 
    diameter approaching \SI{0}{\micro\metre} appear to be undergoing 
    remodelling. Arterial and venular segments present higher diameters 
    ranging up to \SI{34} and \SI{40}{\micro\metre}, respectively.}}
  \label{fig:histo}
\end{figure}

Figure \ref{fig:histo} plots a network diameter histogram (in terms of
total distance covered by vessel segments of a given diameter)
\rev{for models P5B and P6B}. The largest diameter in the network are
\rev{$D_{\textrm{max}} = \SI{34}{\micro \metre}$ and $\SI{40}{\micro
    \metre}$, respectively,} which occur along the retinal vein. The
artery segments have diameters of up to \SI{16}{\micro
  \metre}, with larger diameters closer to the optic
disc. The bulk of the capillary bed has diameters approximately in the
range \SIrange{2}{10}{\micro\metre}, with a reduced amount of vessels
with smaller diameter. These results are substantially lower than the
\emph{in vivo} measurements presented in Table \ref{ta:vels_flows}. In
addition, \rev{a small number of capillaries have diameters
  approaching \SI{0}{\micro\meter}}.  To some extent, these
discrepancies are expected as we are measuring the diameter of the
internal luminal surface with extreme precision (unlike some of the
works cited where only a generic measure of vessel calibre is given),
including in our measurements vessel segments \rev{that appear to be}
undergoing regression (hence in the process of closing up). The
diameter measured for the main arteries are, however, in better
agreement with the \emph{ex vivo} measurements obtained from corrosion
casts by \citet{Ninomiya2006}. \rev{Therefore, we cannot exclude that 
sample preparation and fixation protocols contribute to vessel 
shrinkage.} \rev{Finally, 
  we fit a lognormal probability distribution function to each histogram 
  and use the distribution mode as an estimate of the typical
  capillary diameter (under the assumption that capillaries are the most 
  common vessel type in the
  network). We observe a reduction in the typical capillary
  diameter between day five (\SIlist{5.51;5.78}{\micro\meter}) and six
  (\SIlist{4.44;5.29}{\micro\meter}). More experiments are required in
  order to assess the statistical relevance of these results but the
  implications of a systematic decrease in vessel diameter over time
  are important given that, for a constant flow rate, wall shear stress
  is inversely proportional to the third power of the vessel radius. We
  plan to explore the relationship between changes in geometry and
  haemodynamics as part of a future study.}
%We are
%currently exploring alternative experimental setups that may alleviate
%this problem.
%

In order to ensure that 95\% of the reconstructed network has $\tilde
D\geq 7$, we choose the \rev{voxel sizes in Table \ref{ta:delta_x} for
  the discretisation of each model. \ref{ap:grid_refine} presents a
  grid refinement study aimed at confirming that the choice of voxel
  size leads to spatially converged solutions.}
\begin{table}
\caption{Voxel sizes employed in the discretisation of the different 
  flow models used in this work.}\label{ta:delta_x}
\centering
\begin{tabular}{ccccc}
Model & P5A & P5B & P6A & P6B \\ \hline
$\Delta x$ & \SI{0.5166}{\micro\meter} & \SI{0.5666}{\micro\meter} & \SI{0.5}{\micro\meter} & \SI{0.4166}{\micro\meter} \\ \hline
\end{tabular}
\end{table}

\subsection{Simulations}
\label{sec:real_sim}

Due to its kinetic nature, the LB algorithm
applied to steady flow problems in an initially quiescent domain
requires the system to be advanced in time in order to overcome an
initial transient.
% The main timescale to be captured is the time
% required for momentum to diffuse in the domain
% \begin{equation}
% \tilde t_{\textrm{md}} \approx \frac{\tilde D^2}{\tilde \nu} \approx 125000\;,
% \end{equation}
% where $\tilde D$ is the characteristic length of the system (in our case the
% largest diameter in the network $\tilde D=\tilde D_{\textrm{max}}$) and $\tilde \nu$ is
% taken to be the smallest viscosity predicted by the rheology model
% $\tilde \nu=\tilde \nu_\infty$. Note that from all possible
% choices of $\tilde D$ and $\tilde \nu$, the proposed one represents the worst-case
% scenario.
\rev{In order to monitor convergence, we evaluate the following convergence
criterion at the end of each timestep $t$:
\begin{align}
\label{eq:convergence}
\frac{\max_\vec{r}||\vec{v}(\vec{r})^t - \vec{v}(\vec{r})^{t-1}||}{v_\mathrm{ref}} < \epsilon_\mathrm{tol} = 10^{-6}\;,
\end{align}
where $\vec{v}(\vec{r})^t\coloneqq \vec{v}(\vec{r}, t\Delta t)$ and
$v_\mathrm{ref}$ is a velocity reference value chosen based on the
data summarised in Table \ref{ta:vels_flows}, \emph{i.e.}
$v_\mathrm{ref}=\SI{50}{\milli\metre\per\second}$.  Only when this
condition holds do we consider the
simulation to have reached steady state.}  More efficient methods for
LB initialisation have been proposed (see \emph{e.g.}  \citet{Mei2006,
  Caiazzo2005}) but we will not consider them in the current work
\rev{since our approach remains computationally tractable}.

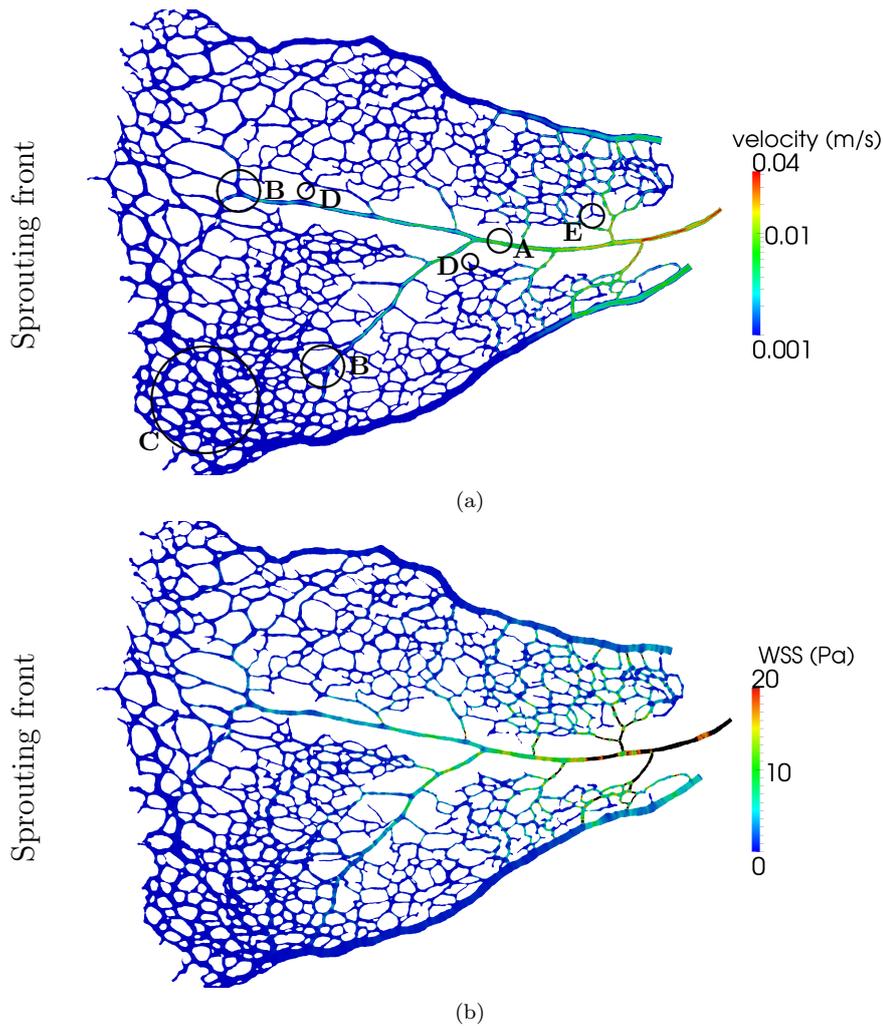
\begin{figure}
  \centering 
  \subfigure[]{
    \label{fig:p5_vel}
    \begin{overpic}[scale=0.1]{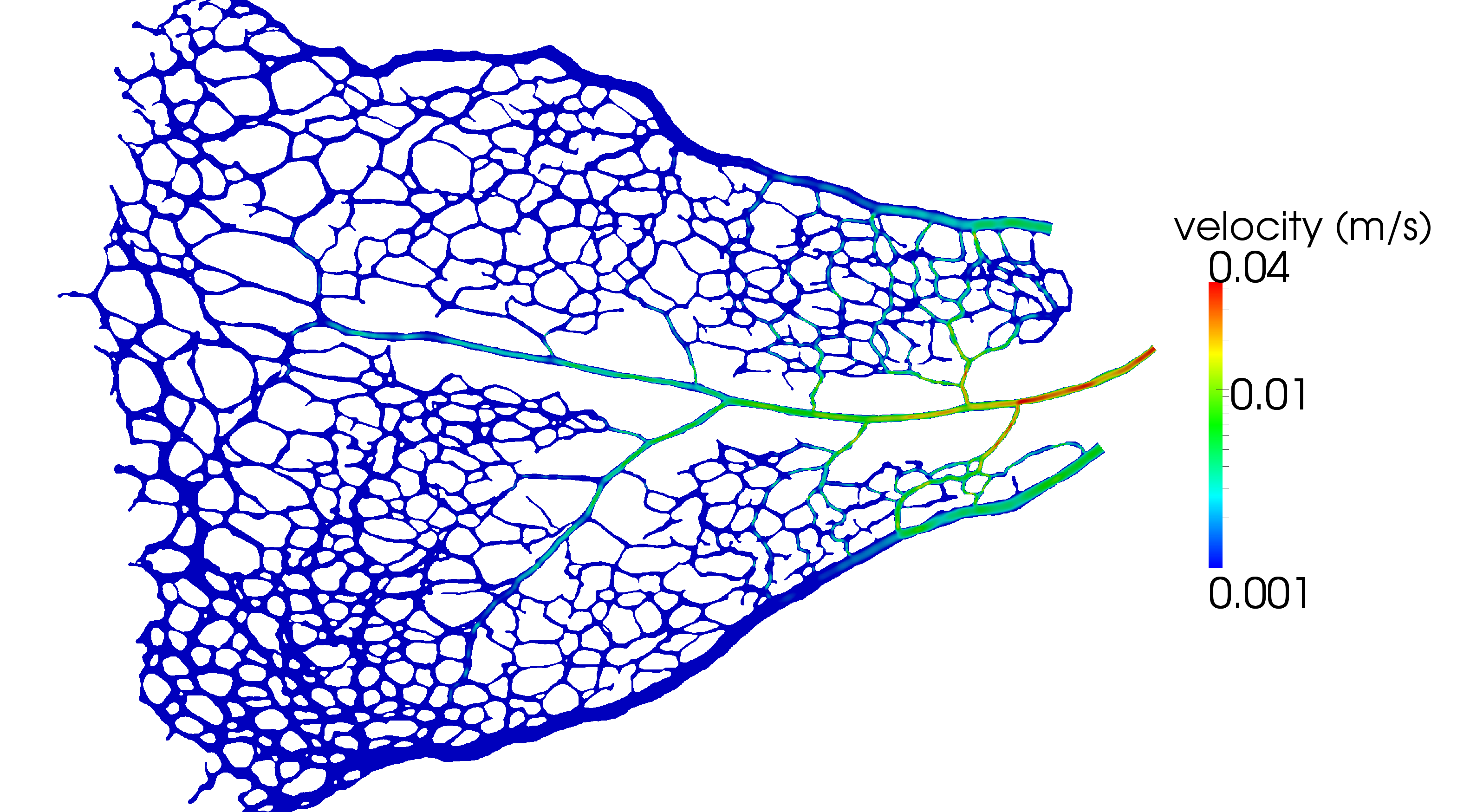}
      \thicklines
      \put(53,28){\color{black}\circle{3}}
      \put(54.5,26){\color{black}{\textbf{A}}}
      \put(22,34){\color{black}\circle{5}}
      \put(25,33){\color{black}{\textbf{B}}}
      \put(32,13){\color{black}\circle{5}}
      \put(35,12){\color{black}{\textbf{B}}}
      \put(18,9){\color{black}\circle{30}}
      \put(10,3){\color{black}{\textbf{C}}}
      \put(30,34){\color{black}\circle{2}}
      \put(31.5,32){\color{black}{\textbf{D}}}
      \put(49.5,25.5){\color{black}\circle{2}}
      \put(45.5,24){\color{black}{\textbf{D}}}
      \put(64,31){\color{black}\circle{3}}
      \put(60.5,28){\color{black}{\textbf{E}}}
      \put(-5,15){\rotatebox{90}{\large Sprouting front}}
    \end{overpic}
  }
  \subfigure[]{
    \label{fig:p5_wss}
    \begin{overpic}[scale=0.1]{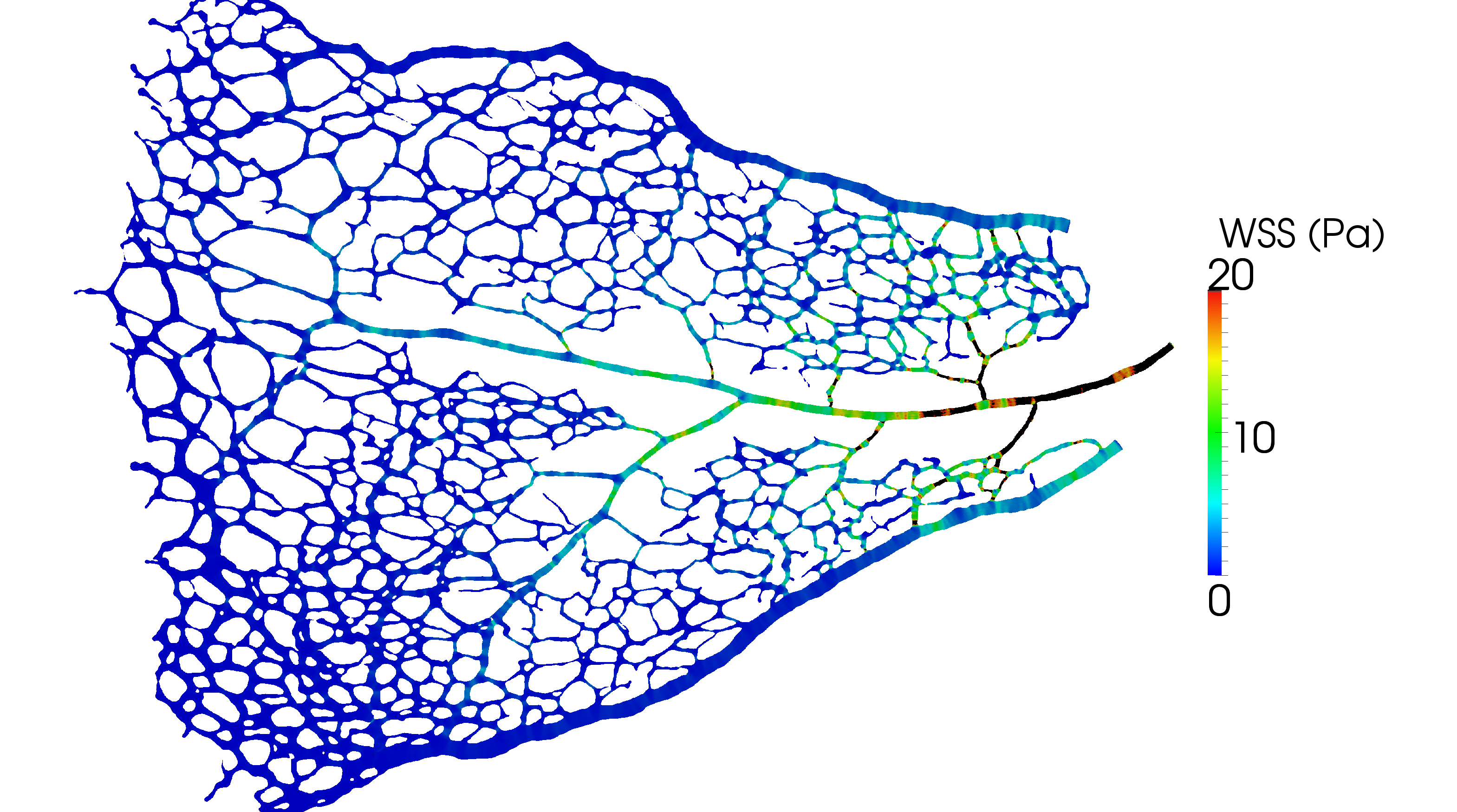}
      \put(-5,15){\rotatebox{90}{\large Sprouting front}}
    \end{overpic}
  }
  \caption{\rev{P5B simulation results: (a) Velocity magnitude plotted on a
    cross-section along the $z=0$ plane. Velocity shows the expected
    parabolic profile across the vessel diameter. Velocity is higher in the artery
    located in the centre of the domain, in particular close to the optic disk. 
    Velocity magnitude quickly decreases as the artery progresses towards the
    sprouting front and it stops being a preferential flow path at the points
    where its identity stops being clearly defined. (b) Wall shear stress (WSS) 
    magnitude plotted on the model surface. Areas of preferential flow tend to 
    experience highest WSS magnitudes. WSS is generally low across the domain 
    except for the arterial segment close to the optic disk and some first order 
    branches. WSS values higher than \SI{20}{\pascal} are considered
    unphysiological and the regions experiencing them are coloured in
    black. Black circles indicate regions of interest referenced in
    the manuscript.}}
  \label{fig:results_p5b}
\end{figure}

\rev{Figure \ref{fig:results_p5b} presents results of a simulation
  with the P5B flow model and the inlet/outlet boundary conditions and
  rheological properties surveyed in Section
  \ref{sec:lit_rev}\ref{sec:ret_haem}. Velocity magnitude is plotted
  at the intersection of the model and the $z = 0$ plane (Figure
  \ref{fig:p5_vel}). Our results show how velocities are larger in the
  central artery (see label A), in particular close to the optic disk. 
  We also note that, as the artery progresses towards the sprouting front, 
  the velocity magnitude decreases rapidly. Furthermore, it stops being a
  preferential flow path at the point where it meets areas of less
  well established vessel identity close to the sprouting front (see
  \emph{e.g.} B regions). Areas with undefined vessel identity are
  correlated with homogeneous velocity distributions (see \emph{e.g.}
  C). There exists evidence of a considerable number of vessels having
  recently regressed along the path of the artery (see \emph{e.g.} D
  branches) and the more developed first order branches (see
  \emph{e.g.} E). The two veins (top and bottom of the images) present
  fewer regressing profiles.}

For a plane with normal $\unitvec{n}$, we define the traction vector
\begin{equation}
\vec{t} = \mat{T}\unitvec{n}\;,
\end{equation}
\emph{i.e.} the force per unit area acting on that plane.  Figure
\ref{fig:p5_wss} plots traction magnitude $||\vec{t}||$ on the model
surface (often referred to as wall shear stress magnitude, WSS). We
observe that areas of preferential flow correlate well with the areas
experiencing larger WSS. \rev{In contrast, vessels in the sprouting front 
are under lower magnitudes of WSS.} The model predicts values of WSS larger
than \SI{20}{\pascal}, which can be deemed unphysiological based on
the microvasculature WSS measurements reported in the literature:
\SI{14}{\pascal} \citep{Popel2005}, \SI{\sim 20}{\pascal}
\citep{Ganesan2010}) or \SI{\sim 13}{\pascal} computed from the values
reported by \citet{Wright2012} (under the assumption of Poiseuille
flow). We believe that the WSS overestimation (mainly occurring at the
central artery and some first order branches) is due to the
vessel shrinkage discussed earlier or other modelling errors.

\begin{figure}
  \centering 
  \subfigure[]{
    \label{fig:realistic_vel}
%    \begin{overpic}[scale=0.1]{images/vel_55mmhg}
    \begin{overpic}[scale=0.1]{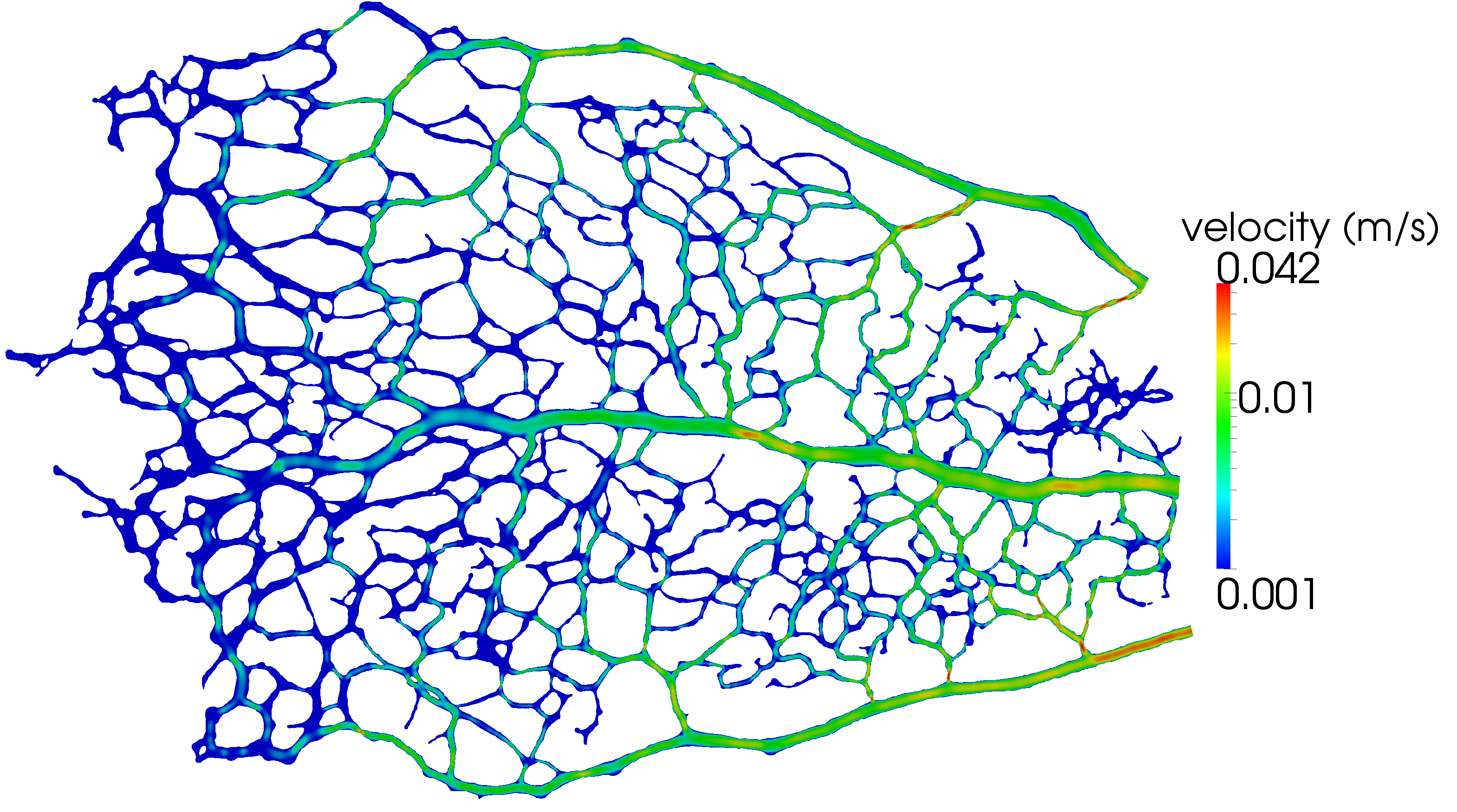}
      \thicklines
      \put(76,28){\color{black}\circle{7}}
      \put(80.5,28){\color{black}{\textbf{C}}}
      \put(67,39.5){\color{black}\circle{3}}
      \put(69,39.5){\color{black}{\textbf{A}}}
      \put(65,19){\color{black}\circle{5}}
      \put(68,19){\color{black}{\textbf{B}}}
      \put(47,20){\color{black}\circle{6}}
      \put(50,22){\color{black}{\textbf{D}}}
      \put(6,32){\color{black}\circle{10}}
      \put(11.5,32.5){\color{black}{\textbf{E}}}
      \put(26,46.5){\color{black}\circle{3}}
      \put(28,44.5){\color{black}{\textbf{F}}}      
      \put(31,42.5){\color{black}\circle{3}}
      \put(33,42.5){\color{black}{\textbf{F}}}
      \put(-5,15){\rotatebox{90}{\large Sprouting front}}
    \end{overpic}
  } \subfigure[]{
    \label{fig:realistic_shear}
%    \begin{overpic}[scale=0.1]{images/wss_55mmhg}
    \begin{overpic}[scale=0.1]{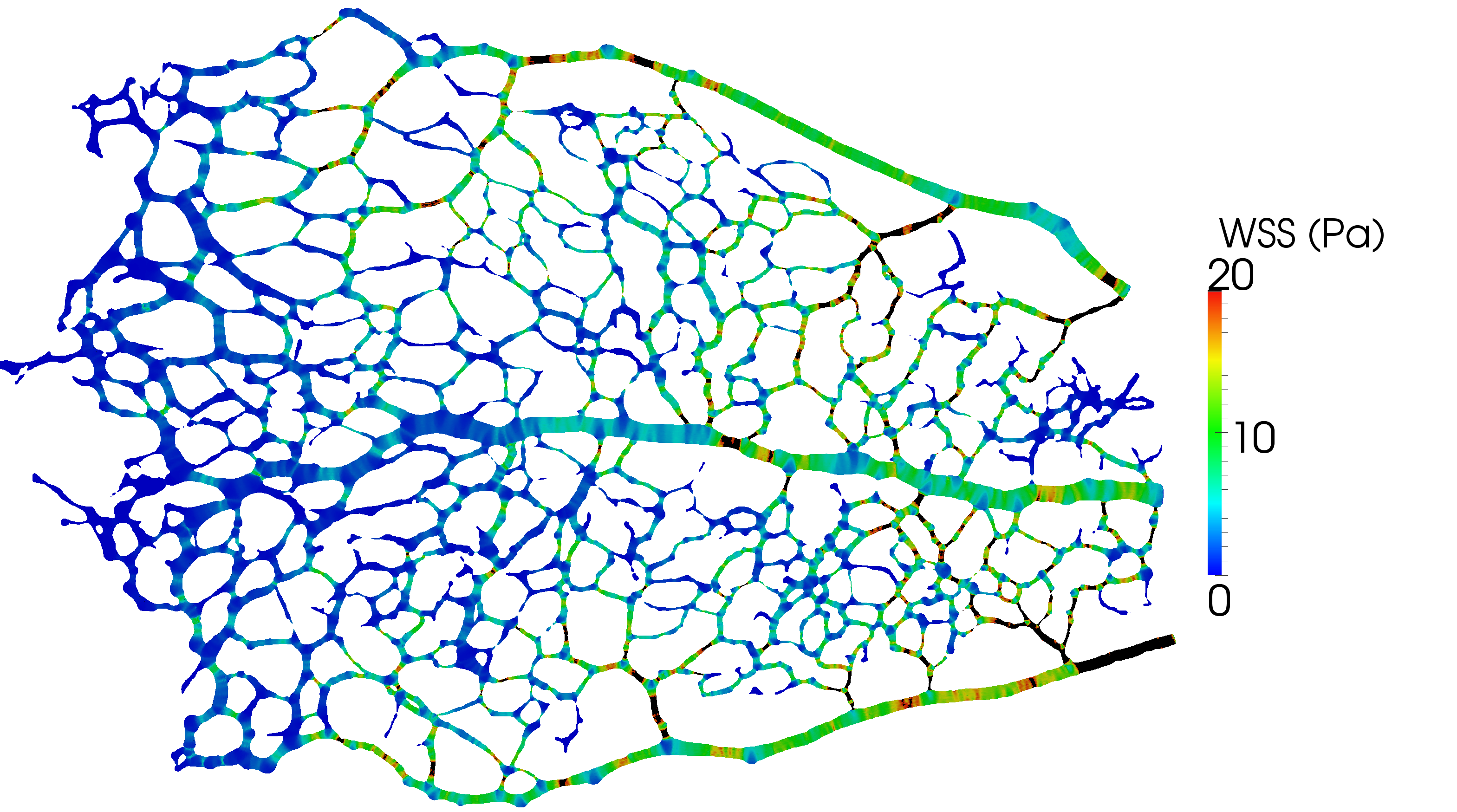}
      \thicklines    
      \put(29.5,52.5){\color{black}\circle{5}}
      \put(-5,15){\rotatebox{90}{\large Sprouting front}}
    \end{overpic}
  }
  \caption{P6A simulation results: (a) Velocity magnitude plotted on a
    cross-section along the $z=0$ plane. Velocity is higher
    in arteries, veins, and segments directly branching from them
    close to the optic disc. Velocity magnitude is smaller in the
    sprouting front. However, vessels of preferential flow already
    exist in the sprouting front; potentially an early indicator of which vessels will survive
    the pruning process. (b) Wall shear stress (WSS) magnitude plotted on the model
    surface. Areas of preferential flow tend to experience highest WSS
    magnitudes. WSS peaks are widely spread across the network. WSS
    magnitude tends to be lower at the junctions and many vessel
    segments present a high-low pattern due to local changes in vessel
    diameter. WSS values higher than \SI{20}{\pascal} are considered
    unphysiological and the regions experiencing them are coloured in
    black. Black circles indicate regions of interest referenced in
    the manuscript.}
\label{fig:realistic_sims}
\end{figure}

Figure \ref{fig:realistic_sims} presents results of a simulation with
the P6A flow model and the inlet/outlet boundary conditions and
rheological properties surveyed in Section
\ref{sec:lit_rev}\ref{sec:ret_haem}.
Figure \ref{fig:realistic_vel} plots velocity magnitude on the
intersection of the model and the $z=0$ plane. First of all, it can be
appreciated how velocities are larger on arteries, veins, and
first-order vessels branching out from them. Highest peak velocities
are around \SI{42}{\milli\metre\per\second} (corresponding to mean
velocities of \SI{21}{\milli\metre\per\second} under Poiseuille flow
assumption) and are in good agreement with the measurements by Wright
and colleagues presented in Table \ref{ta:vels_flows}. Velocity
distribution along a given vessel diameter displays the expected
parabolic profile with zero velocity at the walls. Areas in more
advanced state of pruning (typically closer to the optic disc, see
region B) tend to present larger velocity magnitudes due to a
reduction in vessel density. An exception to this trend is region
C. In this case, we observe a region of very low flow (similar to the
regions found in the less mature vascular plexus towards the
periphery) in an area where pruning should be in a fairly advanced
stage. Two explanations are possible: i) that a recent vessel
regression event has drastically reduced the total flow arriving to
the area which in turn will trigger further vessel regression (similar
to what can be observed in region A) or ii) that a vessel segment
connecting the area with the nearby artery was accidentally removed when
preparing the sample. In contrast to the optic disc region, areas in
the vicinity of the sprouting front experience lower velocity
magnitudes (see region E). Nevertheless, even in this region, we can
already appreciate segments of predominant flow (see \emph{e.g.}  F
regions) rather than a totally homogeneous flow distribution. Taken
together with observations from \citet{Chen2012}, we predict that
these high-flow vessel segments are likely to survive the pruning
process.

Figure \ref{fig:realistic_shear} plots WSS on
the model surface. We observe that areas of preferential flow correlate
well with the areas experiencing larger WSS. However, in this case we
do not see a decrease in WSS with increasing vessel order. WSS peaks
are distributed throughout the domain in agreement with the
observations by \citet{Ganesan2010}. We also observe a complex
distribution of WSS along individual vessel segments, with changes
following local variation in vessel diameter. 

\begin{figure}
  \centering
  \includegraphics[scale=0.1]{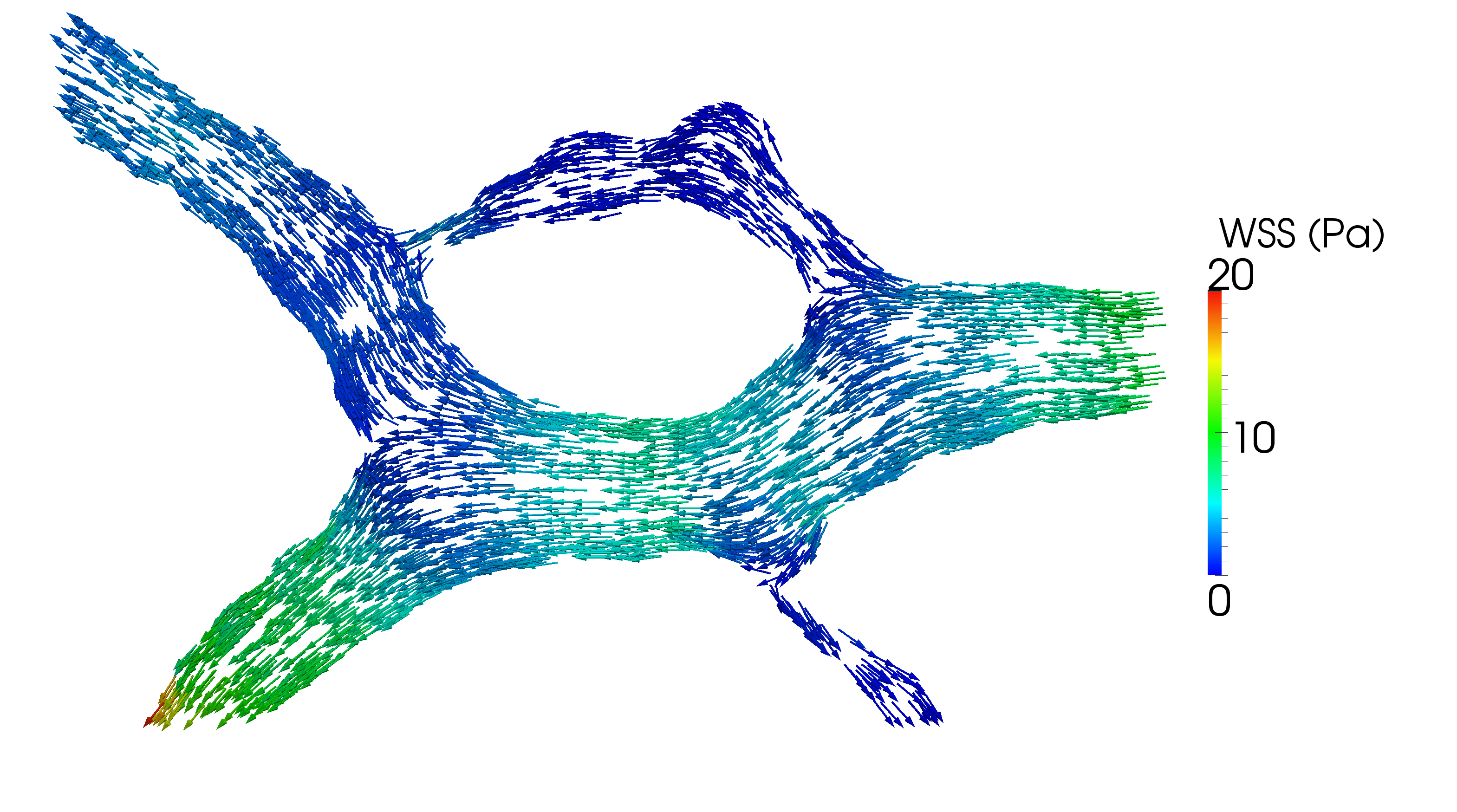}
  \caption{Traction vectors (of constant length and coloured according to
    magnitude) on the luminal surface of the
    region of interest highlighted in Figure
    \ref{fig:realistic_shear}. The loop branch undergoing 
    regression (upper branch) experiences a much lower traction magnitude.}
  \label{fig:shear_stress_vectors}
\end{figure}

Figure \ref{fig:shear_stress_vectors} plots 
$\vec{t}$ on the surface of a subset of the domain (marked with a circle in Figure
\ref{fig:realistic_shear}). Given the redundancy of a loop-like
structure of this type and the distribution of diameters present, it
can be assumed that the upper half of the loop is undergoing
regression. This fact is in good agreement with the distribution of
WSS, of much larger magnitude on the bottom section of the loop and
vessel segments upstream and downstream from it.

In summary, the results presented in this section support the idea
that vessel segments undergoing pruning tend to occur in regions of
low flow (and hence low shear stress). We hypothesise that this
process gradually reduces network density and as a consequence flow
increases in the surviving vessel segments. This in turn prevents
further pruning and contributes to vessel maturation.

% Plexus flow proof of concept simulations. Show that we recover flow
% rates and pressure drops observed experimentally. Present and discuss
% a few wall shear stress maps.

% Remodelled region closer to the optic disc has more homogeneous flow
% distribution. No superfluous vessels, as one would expect.

% Sprouting front was believed to have a much more homegeneous flow
% distribution. We can appreciate predominant branches that likely stay
% after maturation.

% We observe an interesting shear stress pattern with consecutive
% regions of low and high wall shear stress.

\subsection{\rev{Limitations of the study}}
%\todo{Text moved over from Conclusions section}
The main limitations of the study are as follows. First, blood was
modelled as a homogeneous fluid rather than a particle
suspension. This is likely to have an impact on the wall shear stress
computed in small calibre capillaries. \citet{Xiong2010} studied the
changes in haemodynamics induced by the presence of red blood cells
flowing in a simplified model of a microvessel and found up to a 20\%
increase in the shear stress experienced by the luminal wall. Second,
although blood was modelled as a shear-thinning fluid, other
rheological properties such as the F\aa hr\ae us-Lindqvist effect (see
Section \ref{sec:lit_rev}\ref{sec:ret_haem}) were not accounted for.
Third, vessel cross-section was assumed to be circular throughout the
domain due to the lack of spatial information in the z axis. As
previously mentioned, there exists experimental evidence supporting
this assumption in retinal arteries but not in veins
\citep{Feke1989}. This will have an impact on the haemodynamics
recovered.  Also, despite all our efforts when processing retina
samples, we cannot be fully certain that no distortions in the
vascular plexus were introduced.  Next, due to the difficulty of
measuring pressure or flow profiles at the model inlets/outlets
\emph{in vivo} and the absence of suitable data in the literature,
only steady state simulations were performed. We expect flow to be
nearly in phase with pressure given the typical values of Womersley
number (defined as the ratio between oscillatory inertial forces and
viscous forces) encountered in retinal circulation (\SI{\sim0.1}{}
according to \citet{Liu2009}). This makes us confident that flow has
time to fully develop in each cardiac cycle and hence will be well
approximated by an instantaneous pressure gradient. Nevertheless,
there will still be substantial variations in wall shear stress within
any given cardiac cycle. Furthermore, the values of mean arterial
pressure (MAP) and intraocular pressure (IOP) used as inlet and outlet
boundary conditions were obtained from adult animals. In the
supplementary material of this paper (\ref{ap:sens_analy}), 
we perform a sensitivity
analysis of these parameters. Finally, another source of variation in
the predicted haemodynamics are the active and passive mechanical
properties of retinal vessels. At the analysed stage, retinal
arteries are already covered with a smooth muscle layer, which might
contract/relax to control local flow (\emph{i.e.}  autoregulation) and
therefore have an impact in flow patterns in downstream vessels.
% Other limitations:
% Use LB MRT to reduce error

\section{Conclusions}
\label{sec:conclusions}

In the current work, we have presented a software pipeline for the creation
of computational blood flow models based on confocal microscope images
of the microvasculature. The pipeline has been applied to the
development of flow models of the neonatal mouse retinal vasculature
(a common animal model for the study of vascular development). The
different software components used are released under open source
licences.

Using simplified benchmark problems, we have demonstrated the suitability of
the lattice-Boltzmann (LB) algorithm for the simulation of blood flow in sparse and highly
complex vascular networks. Our results indicate that a careful choice
of the LB configuration parameters leads to accurate flow estimates in
channels as narrow as three lattice sites across. Furthermore, we also
showed that the implementation of the no-slip boundary condition
proposed by \citet{Bouzidi2001} produces acceptable estimates of wall
shear stress. We measured errors of $\sim10\%$ and $\sim7\%$ in
channels 7 and 15 lattice sites wide, respectively. Being able to
recover correct haemodynamics even at moderately coarse
discretisations is fundamental to keep the problems under study
computationally tractable.

\rev{In the study reported here, we investigated changes
in haemodynamics during vascular remodelling. Blood flow models were
generated from samples of retinal plexuses obtained at postnatal day
(P) 5 and 6. Our simulations show that, in both cases, velocity and wall
shear stress (WSS) are higher in arteries, veins, and first-order
capillaries closer to the optic disk. However, important differences
in the distribution of velocity and WSS across the domain are observed
when comparing both days (see \emph{e.g.} Figures \ref{fig:p5_wss} and
\ref{fig:realistic_shear}). On the one hand, P5 simulations show a
very homogeneous distribution of velocity and WSS across the capillary
network with moderately high values only in the vicinity of the optic
disk. On the other hand, simulations with the P6 flow model show a
consistently higher and much more spatially complex distribution of
velocity and WSS. Higher values are primarily located in regions in a
more advanced state of remodelling (note \emph{e.g.} the number of
disconnected vessels undergoing regression). In the P6 case, branches
of predominant flow can be also identified in the sprouting front. }

\rev{We also analysed WSS in segments undergoing regression (see
\emph{e.g.} Figure \ref{fig:shear_stress_vectors}) and observed vessel
pruning occurring  in regions of low shear stress. This process
gradually reduces network density (through the removal of redundant
segments) and is likely to lead to an increase in flow in the
surviving vessel segments. We hypothesise that this will contribute to
vessel maturation. Our results support the previously proposed
modulation effect that haemodynamic forces have on developmental
vascular remodelling \citep{Chen2012}.}

\rev{The geometrical analysis of the vascular plexuses leads to two
possible explanations for the increase in velocity and WSS observed
between the P5 and P6 models. First, the increase may be a direct
consequence of the observed decrease in typical capillary diameter
(given the inverse relationship between WSS and the third power of the
vessel radius and assuming that the total flow rate in the retina
remains constant). Second, the progressive reduction in capillary bed
density due to vessel regression may lead to an increase in flow (and
hence WSS) in neighbouring vessels. We believe that both effects may
play complementary roles in order to create the WSS gradients
hypothesised to be behind vessel regression \citep{Chen2012}. Further
experiments are required in order to determine the relative importance
of each of these effects and fully understand how they interact.}

We are currently working on extending the modelling framework to
include tissue mechanics and agent-based cellular modelling. Our goal
is to develop an integrated computational framework for vascular
mechanobiology research. In particular, we are interested in modelling
the interplay between cellular molecular regulation and haemodynamic
forces during vascular remodelling.
% Preliminary results were presented in \cite{BernabeuEMBC13}. % Check
% if the EMBC abstract is out in time.
Finally, the developed methodology should be applicable to
other research domains where small vascular networks can be imaged but
where experimental flow measurements are difficult to obtain.

\section*{Acknowledgements} This work was supported by: Cancer
Research UK; the Lister Institute of Preventive Medicine; the Leducq
Transatlantic Network ARTEMIS; the UK-Israel Initiative BIRAX; EPSRC
grants ``2020 Science'' (\url{http://www.2020science.net/},
EP/I017909/1), ``Large Scale Lattice Boltzmann for Biocolloidal
Systems'' (EP/I034602/1), and ``UK Consortium on Mesoscale Engineering
Sciences (UKCOMES)'' (EP/L00030X/1); and the EC-FP7 projects CRESTA
(\url{http://www.cresta-project.eu/}, grant no. 287703) and MAPPER
(\url{http://www.mapper-project.eu/}, grant no. 261507). CAF is
supported by a Marie Curie Post-doctoral Fellowship of the European
Commission FP7 People framework. This work made use of the HECToR and
ARCHER UK National Supercomputing Services (http://www.archer.ac.uk)
(under EPSRC grants EP/I017909/1 and EP/L00030X/1). The authors
acknowledge the support of the UCL Research Software Development
Service (RSD@UCL) in the completion of this work.

\appendix{\rev{An introduction to the lattice-Boltzmann algorithm}}
\label{ap:lb}

\rev{In the current section we present a brief introduction to the
lattice-Boltzmann (LB) algorithm. The interested reader can refer to
well-cited references such as \citet{Chen1998, Aidun2010} for a more
in-depth presentation and analysis. LB operates at a mesoscopic level,
simulating the evolution of a discrete-velocity approximation to the
one-particle velocity distribution functions of the Boltzmann equation of
kinetic theory, $\{f_i (\vec{r}, t)\}$. Computations are performed on
a regular lattice discretisation of $\bar\Omega_w$, with grid spacing
$\Dx$. The set of velocities $\{\vec{c}_i\}$ is chosen such that the
distances travelled in one timestep ($\Dt$), $\vec{e}_i =
\vec{c}_i\Dt$, are lattice vectors.
When one only wishes to reproduce Navier-Stokes dynamics, the set is
typically a subset of the Moore neighbourhood, including the rest
vector. For 3D simulations, the most commonly used sets have 15, 19
and 27 members. In the current work, we employ the three-dimensional
15 velocity LB lattice (D3Q15).}

\rev{Evolving the distribution functions in time involves two main steps.
The first is known as the collision step, which relaxes the
distributions towards a local equilibrium (the post-collisional
distributions is often denoted as $f_i^\star$):}
\begin{equation}
f_i^\star(\vec{r},t) = f_i(\vec{r},t) + \hat\Omega(f_i(\vec{r},t))\;,
\label{eq:lb-general-relaxation} 
\end{equation} 
\rev{where $\hat\Omega$ is a the collision operator. The second is known as
the streaming step, where $\{f_i^\star\}$ are propagated along the
lattice vectors to new locations in the lattice, defining the
distribution functions at the next timestep:}
\begin{equation} 
f_i(\vec{r} + \vec{c}_i \Dt, t + \Dt) = f_i^\star(\vec{r}, t)\;. 
\end{equation}

\rev{In the current work, we employ the lattice Bhatnagar-Gross-Krook (LBGK) collision
operator, which approximates the collision step as a
relaxation process towards a local equilibrium,}
\begin{equation}\label{eq:lbgk-relaxation}
 \hat\Omega(f_i) = - \frac{(f_i - \feq_i)}{\tau}\Dt\;,
\end{equation}
\rev{where $\tau$ is the relaxation time. This can be shown, through a
Chapman-Enskog expansion (see, e.g., \citet{Ladd2001,Chen1998}), to
reproduce the Navier-Stokes equations in the quasi-incompressible
limit with errors proportional to the lattice Mach number squared. The kinematic viscosity $\nu$ is given by}
\begin{equation}
 \label{eq:lbgk-visc}
 \nu = \cs^2 (\tau - \Dt/2)\;,
\end{equation}
where
\begin{align}
c_s = \frac{\Delta x}{\sqrt{3}\Delta t}\;,
\end{align}
\rev{is the speed of sound in the D3Q15 lattice. Equation \eqref{eq:lbgk-relaxation} can be extended to 
simulate generalised Newtonian flows by locally varying $\tau$ according to \eqref{eq:lbgk-visc} and an
empirical characterisation of the fluid viscosity (\emph{e.g.} Equation \eqref{eq:cy_model}).}

\rev{For the equilibrium distribution, we use a second-order (in velocity
space) approximation to a Maxwellian distribution}
\begin{equation}
\feq_i (\tilde \rho, \tilde{\vec{v}}) = \tilde \rho w_i \left(1 + \frac{\vec{c}_i \cdot \tilde{\vec{v}}}{\tilde{\cs}^2}
+ \frac{ (\vec{c}_i \cdot \tilde{\vec{v}})^2}{2 \tilde{\cs} ^ 4} 
- \frac{\tilde{\vec{v}}\cdot\tilde{\vec{v}}}{2 \tilde{\cs}^2}\right)\;,
\end{equation}
\rev{where the weights $w_i$ and speed of sound $\cs$ depend on the choice
of velocity set. Other choices of $\hat\Omega$ exist. Finally, the
interested reader can refer to previous work by the authors
(\citet{Nash2014} and references therein) on the implementation of
wall and open boundary conditions.}

\rev{The macroscopic density $\rho (\vec{r}, t)$ and velocity
$\vec{v}(\vec{r}, t)$ at a fluid site can be calculated from the
distribution functions by}
\begin{align}
\tilde \rho &= \sum_i f_i\;, \\
\tilde \rho \tilde{\vec{v}} &= \sum_i f_i \vec{c}_i\;.
\end{align}
\rev{The macroscopic pressure is related to the density by the ideal gas
law}
\begin{align}
\tilde P &= \tilde \rho \tilde c^2_s\;.
\end{align}
\rev{A notable characteristic of the LB algorithm is that the shear rate tensor $\mat{S}$ can be computed at any lattice site from local information only. Let $\fneq_i = f_i - \feq_i$ be the non-equilibrium part of the distribution function. It can be shown that (see \emph{e.g.} \citet{Kruger2009}):}
\begin{align}
\tilde{\mat{S}} \approx \frac{-1}{2 \tilde \tau \tilde c_s^2 \tilde \rho}\sum_i \fneq_i \vec{c}_i \vec{c}_i\;.
\end{align}
\rev{The main advantage of this approach is that $\mat{S}$ (and therefore
$\dot{\gamma}$, $\mat{T}$, and $\vec{t}$) can be evaluated locally without the
need of accessing simulation results generated at neighbouring lattice sites
in order to approximate spatial derivatives of the velocity field. This makes
parallel performance independent of the choice of rheology model or the need
of computing wall shear stress (a variable of primary importance in our work).}

\appendix{\rev{Voxel size convergence analysis}}
\label{ap:grid_refine}

\rev{In this section we perform a grid refinement study of the
simulations presented in Section \ref{sec:results}\ref{sec:real_sim}.
We want to assess whether the characterisation of the discretisation
error done in Section \ref{sec:results}\ref{sec:simul_val} with a
simplified domain and rheology model remains valid with more complex
vessel networks and rheology models.}

\rev{In order to confirm that the choice of $\Delta x$ in Table
\ref{ta:delta_x} leads to spatially converged solutions, we generated
a set of increasingly finer discretisations of the P6A flow model. The
finest discretisation that remained computationally tractable for us
had $\Delta x^r = \SI{0.1875}{\micro\metre}$. We used the results of
this simulation as a reference solution and compared it against the
results obtained with $\Delta x^i\in\{1.0,0.5,0.25\}$
\SI{}{\micro\metre}. Note that $\Delta t$ had to be modified according
to \eqref{eq:tau_nu}--\eqref{eq:nu_lattice} in order to keep
$\tilde \tau_{\infty,0}$ constant across all simulations (see Section
\ref{sec:results}\ref{sec:simul_val} for more details). We define the
error in the velocity field for a given discretisation $\Delta x^i$
as}

\begin{align}
\vec{v}^i_\epsilon(\vec{r}^{\Delta x^i}) = \vec{v}^{\Delta x^i}(\vec{r}^{\Delta x^i}, t_\mathrm{conv}) - \vec{v}^{\Delta x^r}(\vec{r}^{\Delta x^r}, t_\mathrm{conv})
\end{align}
\rev{where $t_\mathrm{conv}$ corresponds, in each of the simulations,
to the time where steady flow convergence has been achieved according
to the stop criteria in Equation \ref{eq:convergence} and
$\vec{r}^{\Delta x^r}$ is the lattice site in the reference
discretisation closest to $\vec{r}^{\Delta x^i}$. Finally, we use the
root mean square of the error scaled by the predicted velocity range
as our measure of error}
\begin{align}
\epsilon^2_{u^i} = \frac{\sqrt{\displaystyle \sum_\vec{r} \vec{v}^i_\epsilon \cdot \vec{v}^i_\epsilon}}{\sqrt{N}\;\mathrm{max}_\vec{r}||\vec{v}^{\Delta x^\vec{r}}||}
\end{align}
\rev{where $N$ is the number of lattice sites in the $\Delta x^i$ 
discretisation or a subset of it.}

\begin{figure}
  \centering
  \includegraphics[scale=0.5]{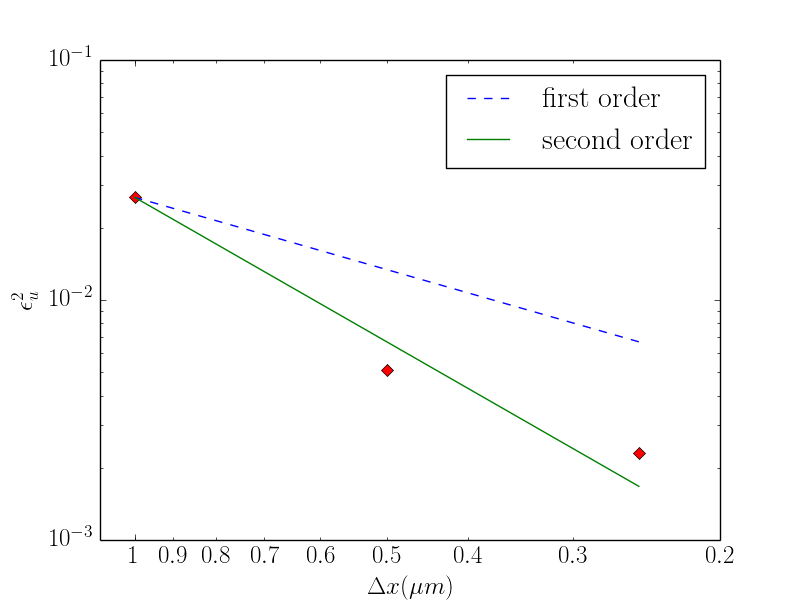}
  \caption{Velocity error residual 
    $\epsilon^2_{u^i}$ on simulations with the P6A flow model 
    discretised with $\Delta x^i =\;$\SIlist{1.0;0.5;0.25}{\micro\metre}
    (red markers). In order to keep the analysis computationally 
    tractable, $\epsilon^2_{u^i}$ is computed with results obtained at 
    the lattice sites located on the $z=0$ plane only (this is the 
    same subset of the results presented in Figure 
    \ref{fig:realistic_sims}). The lines are guides to 
    the eye showing first-order (dashed) and second-order (solid) 
    convergence.} 
  \label{fi:conv_analy}
\end{figure}

\rev{Figure \ref{fi:conv_analy} plots the values of $\epsilon^2_{u^i}$
computed on the subset of the P6A flow model presented in Figure
\ref{fig:realistic_vel} (\emph{i.e.} lattice sites on the $z=0$ plane)
for the choice of $\Delta x$ discussed above. The results show the LB
configuration employed (\emph{i.e.} lattice, collision operator, and
boundary condition implementation) displays second-order convergence
behaviour. WSS (results not presented here) shows similar
convergence trends. These findings are in agreement with results
previously published by the authors \citep{Nash2014}. Finally, our
grid refinement study shows how the choice of voxel sizes in Table
\ref{ta:delta_x} leads to sufficiently spatially converged solutions.
In the case of the P6A flow model, the results generated with  $\Delta
x = \SI{0.5}{\micro\metre}$ have a relative error of only
$\epsilon^2_{u^i}\approx0.005$ when compared with the results obtained
with finest discretisation of the model that remained computationally
tractable.}

\appendix{Inlet/outlet configuration sensitivity analysis}
\label{ap:sens_analy}

In this section we investigate the robustness of the results in
Section \ref{sec:results}\ref{sec:real_sim} with regards to the choice
of inlet and outlet boundary conditions. This is motivated by the fact
that the choice of MAP and IOP in our flow model is based on data
obtained from the literature (see Table \ref{ta:map_iop}) rather than
directly measured from the animals used in the study. First, we want
to investigate whether the general patterns of flow are affected by
moderate changes in OPP (defined as the difference between MAP and
IOP). Second, we want to quantify the relationship between OPP and
blood velocity in the domain.

\begin{figure}
  \centering 
  % \subfigure[\SI{35}{\mmHg}]{
  %   \label{fig:realistic_vel}
  %   \includegraphics[scale=0.08]{images/vel_35mmhg}
  % }
  \subfigure[\SI{45}{\mmHg}]{
    \label{fig:realistic_vel_45}
    \includegraphics[scale=0.08]{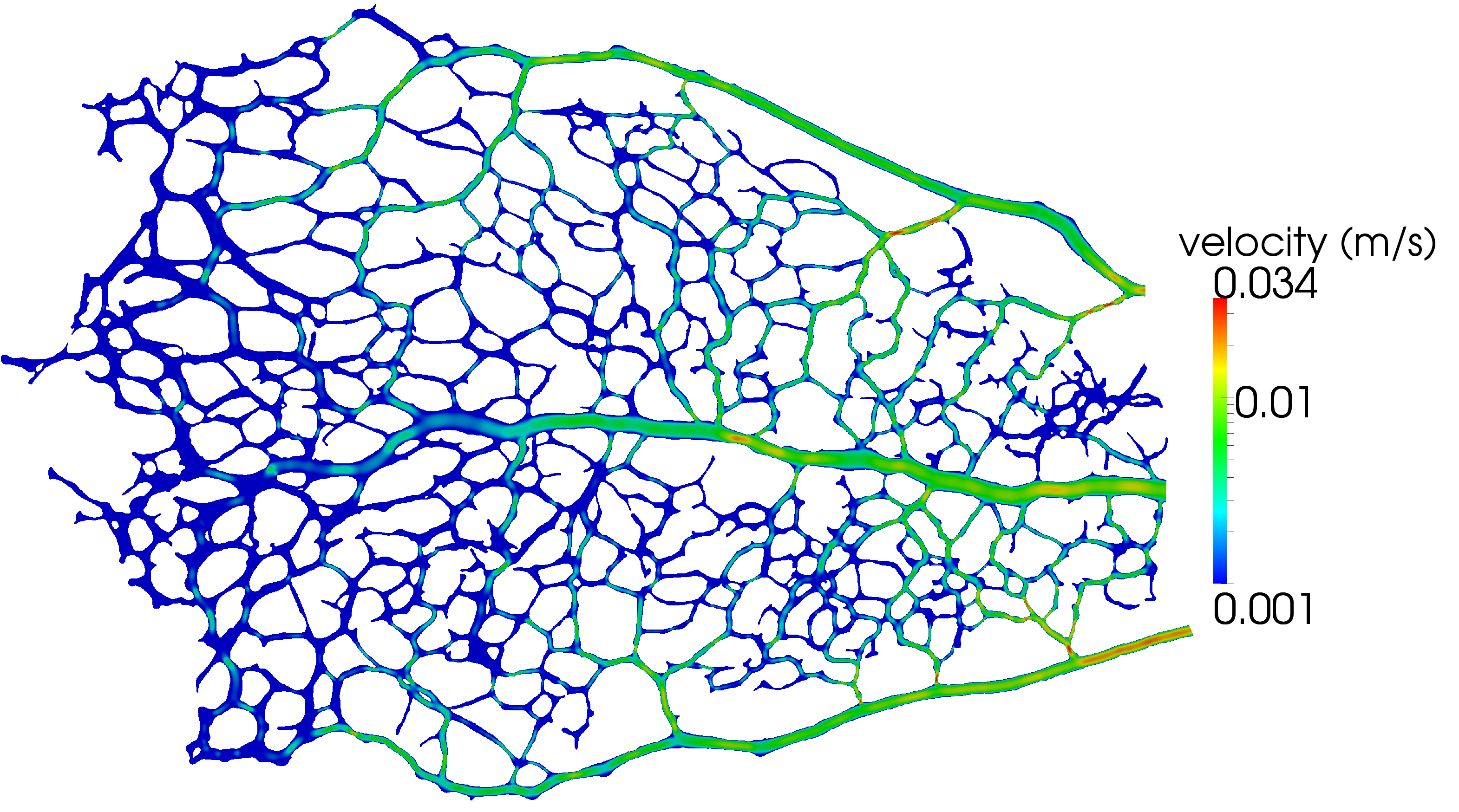}
  } \subfigure[\SI{65}{\mmHg}]{
    \label{fig:realistic_vel_65}
    \includegraphics[scale=0.08]{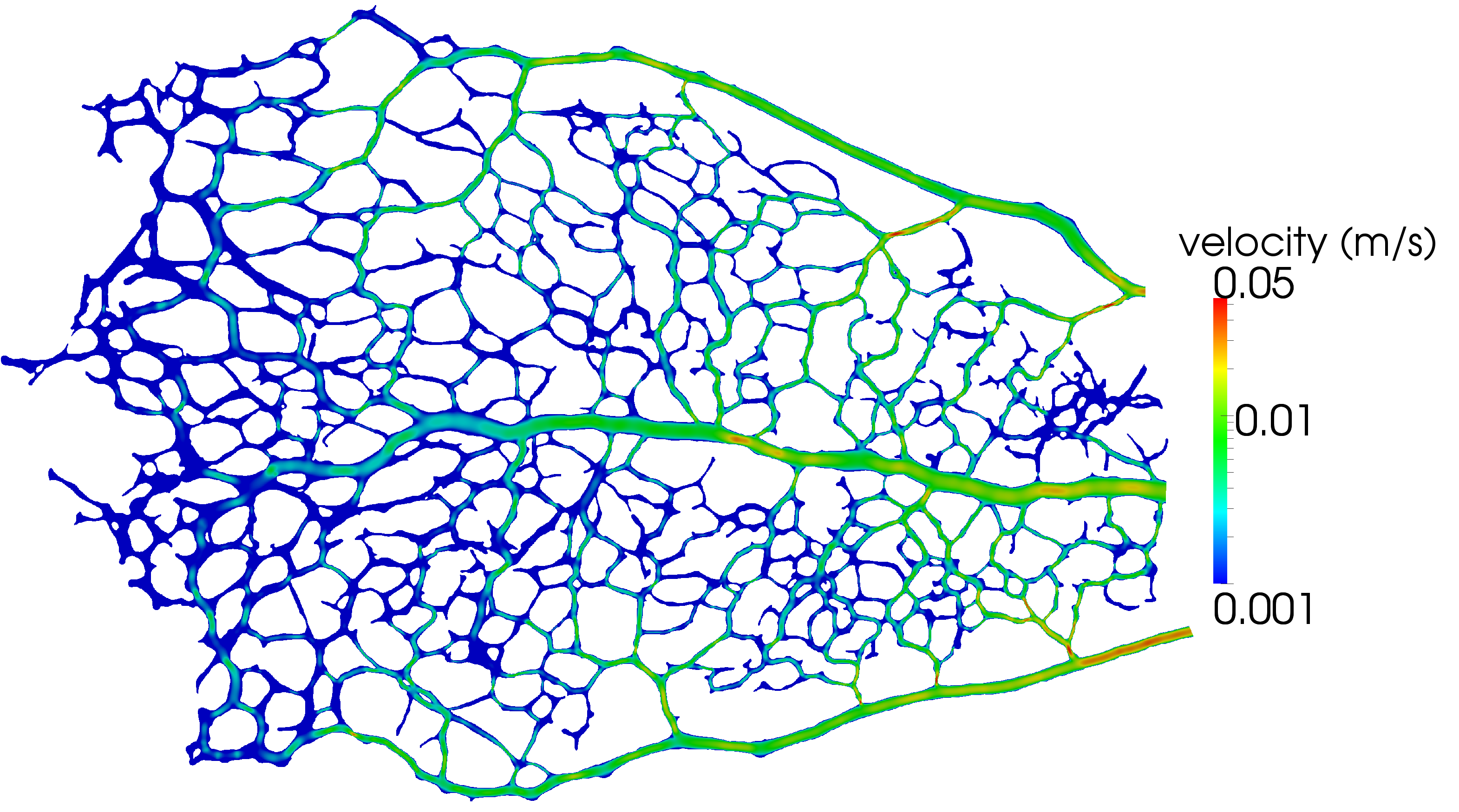}
  }
  \caption{Simulation results: velocity magnitude plotted on a
    cross-section along the $z=0$ plane for OPP values of
    \SI{45}{\mmHg} and \SI{65}{\mmHg}. The logarithmic colour scales
    have been adjusted to range from \SI{1}{\milli\metre\per\second}
    to the largest velocity in the domain. Branches of predominant
    flow and velocity gradients remain fairly constant despite
    moderate changes in OPP when compared to Figure
    \ref{fig:realistic_vel}.}
\label{fig:sens_anal}
\end{figure}

Figure \ref{fig:sens_anal} repeats the visualisation in Figure
\ref{fig:realistic_vel} for simulations with OPP \SI{10}{\mmHg}
smaller and larger. The colour scale has been adjusted to range from
\SI{1}{\milli\metre\per\second} to the largest velocity in the domain
in each case. It can be appreciated how the general patterns of flow
and velocity gradients are greatly preserved from those in Figure
\ref{fig:realistic_sims}. As expected, the absolute values differ.

Table \ref{ta:sens_analysis} presents the largest velocity magnitude
recovered for a wider range of OPP values. It can be seen how peak
velocity increases by approximately \SI{7.6}{\milli\metre\per\second}
for every \SI{10}{\mmHg} increase in OPP. The linear relationship
between pressure difference driving flow in the domain and peak
velocity indicates that flow occurs in the Stokes regime (\emph{i.e.}
inertial forces are small compared to viscous forces). This result is
expected given the typical Reynolds numbers reported in the literature
for microcirculation (\emph{e.g.}  $Re=0.2,0.05, \text{and } 0.0003$
for arterioles, venules, and capillaries, respectively
\citep{Popel2005}).

\begin{table}
  \caption{Peak velocity in the domain as a function of the ocular perfusion
    pressure (OPP).}
\label{ta:sens_analysis}
\centering
\begin{tabular}{cc}
  \\ \hline
  OPP (\SI{}{\mmHg}) & Domain peak velocity (\SI{}{\milli\metre\per\second})  \\ \hline
  25 & 18.97\\
  35 & 26.65\\
  45 & 34.31\\
  55 & 41.96\\
  65 & 49.61\\
  \hline
\end{tabular}
\end{table}

We conclude that the inlet/outlet boundary conditions play a secondary
role in determining the flow patterns in the network, hence no
modulation in the WSS experienced by the endothelium is expected via
changes in OPP (this does not include changes due to autoregulation of
the vessels themselves). Changes in vessel network geometry are
expected to be the main drivers behind haemodynamic reorganisation
during development. This observation supports the idea that regression
of a given vessel segment will affect the flow patterns in nearby
segments, leading to potential increases in WSS that may contribute to
vessel maturation.

\bibliographystyle{apalike}
\bibliography{library.bib}

\begin{thebibliography}{}

\bibitem[Aidun and Clausen, 2010]{Aidun2010}
Aidun, C.~K. and Clausen, J.~R. (2010).
\newblock {Lattice-Boltzmann Method for Complex Flows}.
\newblock {\em Annual Review of Fluid Mechanics}, 42(1):439--472.

\bibitem[Alm and Bill, 1973]{Alm1973}
Alm, A. and Bill, A. (1973).
\newblock {Ocular and optic nerve blood flow at normal and increased
  intraocular pressures in monkeys (Macaca irus): a study with radioactively
  labelled microspheres including flow determinations in brain and some other
  tissues}.
\newblock {\em Experimental Eye Research}, 15(1):15--29.

\bibitem[Antiga et~al., 2008]{Antiga2008}
Antiga, L., Piccinelli, M., Botti, L., Ene-Iordache, B., Remuzzi, A., and
  Steinman, D.~A. (2008).
\newblock {An image-based modeling framework for patient-specific computational
  hemodynamics.}
\newblock {\em Medical \& biological engineering \& computing},
  46(11):1097--112.

\bibitem[Barber et~al., 1996]{Barber1996}
Barber, C.~B., Dobkin, D.~P., and Huhdanpaa, H. (1996).
\newblock {The quickhull algorithm for convex hulls}.
\newblock {\em ACM Transactions on Mathematical Software}, 22(4):469--483.

\bibitem[Bentley et~al., 2013]{Bentley2013}
Bentley, K., Jones, M., and Cruys, B. (2013).
\newblock {Predicting the future: Towards symbiotic computational and
  experimental angiogenesis research.}
\newblock {\em Experimental cell research}, 319(9):1240--6.

\bibitem[Bernabeu et~al., 2013]{Bernabeu2013Rheology}
Bernabeu, M.~O., Nash, R.~W., Groen, D., Carver, H.~B., Hetherington, J.,
  Kr\"{u}ger, T., and Coveney, P.~V. (2013).
\newblock {Impact of blood rheology on wall shear stress in a model of the
  middle cerebral artery}.
\newblock {\em Interface Focus}, 3(2):20120094.

\bibitem[Bouzidi et~al., 2001]{Bouzidi2001}
Bouzidi, M., Firdaouss, M., and Lallemand, P. (2001).
\newblock {Momentum transfer of a Boltzmann-lattice fluid with boundaries}.
\newblock {\em Physics of Fluids}, 13(11):3452.

\bibitem[Boyd et~al., 2007]{Boyd2007}
Boyd, J., Buick, J.~M., and Green, S. (2007).
\newblock {Analysis of the Casson and Carreau-Yasuda non-Newtonian blood models
  in steady and oscillatory flows using the lattice Boltzmann method}.
\newblock {\em Physics of Fluids}, 19(9):093103.

\bibitem[Brown et~al., 2005]{Brown2005}
Brown, A.~S., Leamen, L., Cucevic, V., and Foster, F.~S. (2005).
\newblock {Quantitation of hemodynamic function during developmental vascular
  regression in the mouse eye.}
\newblock {\em Investigative ophthalmology \& visual science}, 46(7):2231--7.

\bibitem[Caiazzo, 2005]{Caiazzo2005}
Caiazzo, A. (2005).
\newblock {Analysis of Lattice Boltzmann Initialization Routines}.
\newblock {\em Journal of Statistical Physics}, 121(1-2):37--48.

\bibitem[Chen et~al., 2012]{Chen2012}
Chen, Q., Jiang, L., Li, C., Hu, D., Bu, J.-w., Cai, D., and Du, J.-l. (2012).
\newblock {Haemodynamics-driven developmental pruning of brain vasculature in
  zebrafish.}
\newblock {\em PLoS Biology}, 10(8):e1001374.

\bibitem[Chen and Doolen, 1998]{Chen1998}
Chen, S. and Doolen, G.~D. (1998).
\newblock {Lattice Boltzmann method for fluid flows}.
\newblock {\em Annual Review of Fluid Mechanics}, 30(1):329--364.

\bibitem[Chien, 1970]{Chien1970}
Chien, S. (1970).
\newblock {Shear Dependence of Effective Cell Volume as a Determinant of Blood
  Viscosity}.
\newblock {\em Science}, 168(3934):977--979.

\bibitem[Feke et~al., 1989]{Feke1989}
Feke, G.~T., Tagawa, H., Deupree, D.~M., Goger, D.~G., Sebag, J., and Weiter,
  J.~J. (1989).
\newblock {Blood flow in the normal human retina.}
\newblock {\em Investigative ophthalmology \& visual science}, 30(1):58--65.

\bibitem[Formaggia et~al., 2009]{FormaggiaBook}
Formaggia, L., Quarteroni, A., and Veneziani, A. (2009).
\newblock {\em {Cardiovascular Mathematics. Modeling and simulation of the
  circulatory system.}}
\newblock Springer-Verlag Italia, Milano.

\bibitem[Franco et~al., 2008]{Franco2008}
Franco, C.~A., Mericskay, M., Parlakian, A., Gary-Bobo, G., Gao-Li, J., Paulin,
  D., Gustafsson, E., and Li, Z. (2008).
\newblock {Serum response factor is required for sprouting angiogenesis and
  vascular integrity.}
\newblock {\em Developmental cell}, 15(3):448--61.

\bibitem[Fruttiger, 2007]{Fruttiger2007}
Fruttiger, M. (2007).
\newblock {Development of the retinal vasculature.}
\newblock {\em Angiogenesis}, 10(2):77--88.

\bibitem[Ganesan et~al., 2010]{Ganesan2010}
Ganesan, P., He, S., and Xu, H. (2010).
\newblock {Development of an image-based network model of retinal vasculature.}
\newblock {\em Annals of biomedical engineering}, 38(4):1566--85.

\bibitem[Gariano and Gardner, 2005]{Gariano2005}
Gariano, R.~F. and Gardner, T.~W. (2005).
\newblock {Retinal angiogenesis in development and disease.}
\newblock {\em Nature}, 438(7070):960--6.

\bibitem[Geudens and Gerhardt, 2011]{Geudens2011}
Geudens, I. and Gerhardt, H. (2011).
\newblock {Coordinating cell behaviour during blood vessel formation.}
\newblock {\em Development}, 138(21):4569--83.

\bibitem[Groen et~al., 2013]{Groen2013Performance}
Groen, D., Hetherington, J., Carver, H.~B., Nash, R.~W., Bernabeu, M.~O., and
  Coveney, P.~V. (2013).
\newblock {Analysing and modelling the performance of the HemeLB
  lattice-Boltzmann simulation environment}.
\newblock {\em Journal of Computational Science}, 4(5):412--422.

\bibitem[Hardy et~al., 1994]{Hardy1994}
Hardy, P., Abran, D., Li, D.~Y., Fernandez, H., Varma, D.~R., and Chemtob, S.
  (1994).
\newblock {Free radicals in retinal and choroidal blood flow autoregulation in
  the piglet: interaction with prostaglandins.}
\newblock {\em Investigative ophthalmology \& visual science}, 35(2):580--91.

\bibitem[Heywood et~al., 1996]{Heywood1996}
Heywood, J.~G., Rannacher, R., and Turek, S. (1996).
\newblock {Artificial boundaries and flux and pressure conditions for the
  incompressible Navier-Stokes equations}.
\newblock {\em International Journal for Numerical Methods in Fluids},
  22(5):325--352.

\bibitem[Jones et~al., 2006]{Jones2006}
Jones, E. a.~V., le~Noble, F., and Eichmann, A. (2006).
\newblock {What determines blood vessel structure? Genetic prespecification vs.
  hemodynamics.}
\newblock {\em Physiology}, 21:388--95.

\bibitem[Kiel, 2010]{Kiel2010}
Kiel, J.~W. (2010).
\newblock {\em {The Ocular Circulation}}.
\newblock Morgan \& Claypool Life Sciences.

\bibitem[Kr\"{u}ger et~al., 2009]{Kruger2009}
Kr\"{u}ger, T., Varnik, F., and Raabe, D. (2009).
\newblock {Shear stress in lattice Boltzmann simulations}.
\newblock {\em Physical Review E}, 79(4):046704.

\bibitem[Ladd and Verberg, 2001]{Ladd2001}
Ladd, A. and Verberg, R. (2001).
\newblock {Lattice-Boltzmann simulations of particle-fluid suspensions}.
\newblock {\em Journal of Statistical Physics}, 104(September):1191--1251.

\bibitem[Ladd, 1994]{Ladd1994}
Ladd, A. J.~C. (1994).
\newblock {Numerical simulations of particulate suspensions via a discretized
  Boltzmann equation. Part 1. Theoretical foundation}.
\newblock {\em Journal of Fluid Mechanics}, 271:285.

\bibitem[L\"{a}tt and Chopard, 2008]{Latt2008}
L\"{a}tt, J. and Chopard, B. (2008).
\newblock {Straight velocity boundaries in the lattice Boltzmann method}.
\newblock {\em Physical Review E}, 77(5):056703.

\bibitem[Lawson and Weinstein, 2002]{Lawson2002}
Lawson, N.~D. and Weinstein, B.~M. (2002).
\newblock {Arteries and veins: making a difference with zebrafish.}
\newblock {\em Nature reviews. Genetics}, 3(9):674--82.

\bibitem[Liu et~al., 2009]{Liu2009}
Liu, D., Wood, N.~B., Witt, N., Hughes, A.~D., Thom, S.~A., and Xu, X.~Y.
  (2009).
\newblock {Computational Analysis of Oxygen Transport in the Retinal Arterial
  Network}.
\newblock {\em Current Eye Research}, 34(11):945--956.

\bibitem[Lobov et~al., 2005]{Lobov2005}
Lobov, I.~B., Rao, S., Carroll, T.~J., Vallance, J.~E., Ito, M., Ondr, J.~K.,
  Kurup, S., Glass, D.~a., Patel, M.~S., Shu, W., Morrisey, E.~E., McMahon,
  A.~P., Karsenty, G., and Lang, R.~a. (2005).
\newblock {WNT7b mediates macrophage-induced programmed cell death in
  patterning of the vasculature.}
\newblock {\em Nature}, 437(7057):417--21.

\bibitem[Mazzeo and Coveney, 2008]{Mazzeo2008}
Mazzeo, M.~D. and Coveney, P.~V. (2008).
\newblock {HemeLB: A high performance parallel lattice-Boltzmann code for large
  scale fluid flow in complex geometries}.
\newblock {\em Computer Physics Communications}, 178(12):894--914.

\bibitem[Mei et~al., 2006]{Mei2006}
Mei, R., Luo, L.-S., Lallemand, P., and D'Humi\`{e}res, D. (2006).
\newblock {Consistent initial conditions for lattice Boltzmann simulations}.
\newblock {\em Computers \& Fluids}, 35(8-9):855--862.

\bibitem[Nagaoka and Yoshida, 2006]{Nagaoka2006}
Nagaoka, T. and Yoshida, A. (2006).
\newblock {Noninvasive evaluation of wall shear stress on retinal
  microcirculation in humans.}
\newblock {\em Investigative ophthalmology \& visual science}, 47(3):1113--9.

\bibitem[Nash et~al., 2014]{Nash2014}
Nash, R.~W., Carver, H.~B., Bernabeu, M.~O., Hetherington, J., Groen, D.,
  Kr\"{u}ger, T., and Coveney, P.~V. (2014).
\newblock {Choice of boundary condition and collision operator for
  lattice-Boltzmann simulation of moderate Reynolds number flow in complex
  domains}.
\newblock {\em Physical Review E}, 89:023303.

\bibitem[Ninomiya and Inomata, 2006]{Ninomiya2006}
Ninomiya, H. and Inomata, T. (2006).
\newblock {Microvasculature of the mouse eye: Scanning electron microscopy of
  vascular corrosion casts}.
\newblock {\em Journal of Experimental Animal Science}, 43(3):149--159.

\bibitem[Paques, 2003]{Paques2003}
Paques, M. (2003).
\newblock {Structural and Hemodynamic Analysis of the Mouse Retinal
  Microcirculation}.
\newblock {\em Investigative Ophthalmology \& Visual Science},
  44(11):4960--4967.

\bibitem[Popel and Johnson, 2005]{Popel2005}
Popel, A.~S. and Johnson, P.~C. (2005).
\newblock {Microcirculation and Hemorheology.}
\newblock {\em Annual review of fluid mechanics}, 37:43--69.

\bibitem[Potente et~al., 2011]{Potente2011}
Potente, M., Gerhardt, H., and Carmeliet, P. (2011).
\newblock {Basic and therapeutic aspects of angiogenesis.}
\newblock {\em Cell}, 146(6):873--87.

\bibitem[Pries et~al., 1996]{Pries1996}
Pries, A.~R., Secomb, T.~W., and Gaehtgens, P. (1996).
\newblock {Biophysical aspects of blood flow in the microvasculature.}
\newblock {\em Cardiovascular research}, 32(4):654--67.

\bibitem[Schroeder et~al., 2003]{Schroeder2003}
Schroeder, W.~J., Martin, K., and Lorensen, W.~E. (2003).
\newblock {The Visualization Toolkit: An Object-Oriented Approach to 3D
  Graphics, Third Edition}.

\bibitem[Siekmann and Lawson, 2007]{Siekmann2007}
Siekmann, A.~F. and Lawson, N.~D. (2007).
\newblock {Notch signalling limits angiogenic cell behaviour in developing
  zebrafish arteries.}
\newblock {\em Nature}, 445(7129):781--4.

\bibitem[Siggers and Waters, 2005]{Siggers2005}
Siggers, J.~H. and Waters, S.~L. (2005).
\newblock {Steady flows in pipes with finite curvature}.
\newblock {\em Physics of Fluids}, 17(7):077102.

\bibitem[Stahl et~al., 2010]{Stahl2010}
Stahl, B., Chopard, B., and Latt, J. (2010).
\newblock {Measurements of wall shear stress with the lattice Boltzmann method
  and staircase approximation of boundaries}.
\newblock {\em Computers \& Fluids}, 39(9):1625--1633.

\bibitem[Uemura et~al., 2006]{Uemura2006}
Uemura, A., Kusuhara, S., Katsuta, H., and Nishikawa, S.-I. (2006).
\newblock {Angiogenesis in the mouse retina: a model system for experimental
  manipulation.}
\newblock {\em Experimental cell research}, 312(5):676--83.

\bibitem[Vogel et~al., 2003]{Vogel2003}
Vogel, J., Kiessling, I., Heinicke, K., Stallmach, T., Ossent, P., Vogel, O.,
  Aulmann, M., Frietsch, T., Schmid-Sch\"{o}nbein, H., Kuschinsky, W., and
  Gassmann, M. (2003).
\newblock {Transgenic mice overexpressing erythropoietin adapt to excessive
  erythrocytosis by regulating blood viscosity.}
\newblock {\em Blood}, 102(6):2278--84.

\bibitem[Wang et~al., 2007]{Wang2007}
Wang, Y., Bower, B.~a., Izatt, J.~a., Tan, O., and Huang, D. (2007).
\newblock {In vivo total retinal blood flow measurement by Fourier domain
  Doppler optical coherence tomography.}
\newblock {\em Journal of biomedical optics}, 12(4):041215.

\bibitem[Wang et~al., 2009]{Wang2009}
Wang, Y., Fawzi, A., Tan, O., Gil-Flamer, J., and Huang, D. (2009).
\newblock {Retinal blood flow detection in diabetic patients by Doppler Fourier
  domain optical coherence tomography}.
\newblock {\em Optics Express}, 17(5):4061.

\bibitem[Watson et~al., 2012]{Watson2012}
Watson, M.~G., McDougall, S.~R., Chaplain, M. a.~J., Devlin, a.~H., and
  Mitchell, C.~a. (2012).
\newblock {Dynamics of angiogenesis during murine retinal development: a
  coupled in vivo and in silico study.}
\newblock {\em Journal of the Royal Society, Interface / the Royal Society},
  9(74):2351--64.

\bibitem[White et~al., 2003]{White2003}
White, B., Pierce, M., Nassif, N., Cense, B., Park, B., Tearney, G., Bouma, B.,
  Chen, T., and de~Boer, J. (2003).
\newblock {In vivo dynamic human retinal blood flow imaging using
  ultra-high-speed spectral domain optical coherence tomography.}
\newblock {\em Optics express}, 11(25):3490--7.

\bibitem[Windberger et~al., 2003]{Windberger2003}
Windberger, U., Bartholovitsch, A., Plasenzotti, R., Korak, K., and Heinze, G.
  (2003).
\newblock {Whole blood viscosity, plasma viscosity and erythrocyte aggregation
  in nine mammalian species: reference values and comparison of data}.
\newblock {\em Exp Physiol}, 88(3):431--440.

\bibitem[Windberger et~al., 2005]{Windberger2005}
Windberger, U., Grohmann, K., Goll, a., Plasenzotti, R., and Losert, U. (2005).
\newblock {Fetal and juvenile animal hemorheology.}
\newblock {\em Clinical hemorheology and microcirculation}, 32(3):191--7.

\bibitem[Wright and Harris, 2008]{Wright2008}
Wright, W.~S. and Harris, N.~R. (2008).
\newblock {Ozagrel attenuates early streptozotocin-induced constriction of
  arterioles in the mouse retina.}
\newblock {\em Experimental eye research}, 86(3):528--36.

\bibitem[Wright et~al., 2009]{Wright2009}
Wright, W.~S., Messina, J.~E., and Harris, N.~R. (2009).
\newblock {Attenuation of diabetes-induced retinal vasoconstriction by a
  thromboxane receptor antagonist.}
\newblock {\em Experimental eye research}, 88(1):106--12.

\bibitem[Wright et~al., 2012]{Wright2012}
Wright, W.~S., Yadav, A.~S., McElhatten, R.~M., and Harris, N.~R. (2012).
\newblock {Retinal blood flow abnormalities following six months of
  hyperglycemia in the Ins2(Akita) mouse.}
\newblock {\em Experimental eye research}, 98:9--15.

\bibitem[Xiong and Zhang, 2010]{Xiong2010}
Xiong, W. and Zhang, J. (2010).
\newblock {Shear stress variation induced by red blood cell motion in
  microvessel.}
\newblock {\em Annals of biomedical engineering}, 38(8):2649--59.

\bibitem[Yazdanfar, 2003]{Yazdanfar2003}
Yazdanfar, S. (2003).
\newblock {In Vivo Imaging of Human Retinal Flow Dynamics by Color Doppler
  Optical Coherence Tomography}.
\newblock {\em Archives of Ophthalmology}, 121(2):235.

\bibitem[Zhi et~al., 2011]{Zhi2011}
Zhi, Z., Cepurna, W., Johnson, E., Shen, T., Morrison, J., and Wang, R.~K.
  (2011).
\newblock {Volumetric and quantitative imaging of retinal blood flow in rats
  with optical microangiography.}
\newblock {\em Biomedical optics express}, 2(3):579--91.

\bibitem[Zhi et~al., 2012]{Zhi2012}
Zhi, Z., Yin, X., Dziennis, S., Wietecha, T., Hudkins, K.~L., Alpers, C.~E.,
  and Wang, R.~K. (2012).
\newblock {Optical microangiography of retina and choroid and measurement of
  total retinal blood flow in mice.}
\newblock {\em Biomedical optics express}, 3(11):2976--86.

\bibitem[Ziegler, 1993]{Ziegler1993}
Ziegler, D.~P. (1993).
\newblock {Boundary conditions for lattice Boltzmann simulations}.
\newblock {\em Journal of Statistical Physics}, 71(5-6):1171--1177.

\end{thebibliography}

\end{document}